# Fine-tuning the Microstructure and Photophysical Characteristics of Fluorescent Conjugated Copolymers Using Photoalignment and Liquid-crystalline Ordering

Yuping Shi,* Katharina Landfester, and Stephen M. Morris*

Dr. Y. Shi, Professor K. Landfester
Max Planck Institute for Polymer Research, Mainz 55128, Germany
Email: shiy@mpip-mainz.mpg.de

Dr. Y. Shi, Professor S. M. Morris
Department of Engineering Science, University of Oxford, Parks Road, Oxford, OX1 3PJ, UK
Email: stephen.morris@eng.ox.ac.uk

## Abstract

Replicating the microstructure and near-unity excitation energy transfer efficiency in natural light-harvesting complexes (LHCs) remains a major challenge for synthetic energy-harvesting devices. Biological photosynthesis can spontaneously regulate the active ensembles of involved energy absorbing and funnelling chlorophyll-containing proteins in response to fluctuating sunlight. Here we utilize liquid crystalline (LC) ordering to fine-tune the polymer packing and photophysical properties in liquid crystalline conjugated polymer (LCCP) films for LHC biomimicry and optimizing photoluminescence quantum efficiency (PLQE). We show that the long-range orientational ordering present in a LC phase of poly(9,9-dioctylfluorene-co-benzothiadiazole) (F8BT) stabilizes a small fraction of randomly-oriented F8BT nanocrystals dispersed in an amorphous matrix of disordered F8BT chains, hence resembling a self-doped host-guest system whereby excitation energy funnelling and PLQE are reinforced significantly by three-dimensional donor-to-acceptor Förster resonance energy transfer (FRET) and dominant intrachain emission in the nano-crystalline acceptor. Furthermore, the photoalignment of nematic F8BT layers is combined to fabricate long-sought large-area-extended monodomains which exhibit >60% crystallinity and ~20 nm-long interchain packing order, whilst also promoting linearly polarized emission, a new band-edge absorption species, and an extra emissive interchain excited state. Our micro-PL spectral results support the feasibility of making use of self-doped F8BT nematic films for bio-mimicry of certain structural basis and light-harvesting properties of naturally occurring LHCs.

# 1. Introduction

Nature is a constant and unparalleled source of inspiration for microstructural design and many-body coupling optimization. Higher plants and many algae are able to make efficient use of physical-chemical tuning of protein-templated pigment chromophores within a light-harvesting complex (LHC) in order to fully support and spontaneously regulate their photosynthetic activities.[1] Absorbed photon energy is spatially funnelled at ultra-low energy losses among LHCs and ultimately arrives at an energetically expensive reaction centre for charge separation and chemical transformations.[2,3] Consecutive FRET or even quantum coherent energy transfer between these chemically similar yet differently-sized LHCs are unique in terms of their ability to enable near-unity efficiencies of excitation energy funnelling from multiple minor chromophoric LHCs (LH2) to a structurally identical but larger-sized major acceptor complex (LH1).[4,5] These multichromophoric LH1 and LH2 complexes in ubiquitous photosystems have evolved naturally to take an elegant ring-shaped packing microstructure with precisely templated chromophoric molecules by a protein matrix,[1-5] which could serve as a useful guide for the design and manufacture of man-made light-harvesting and energy-conversion devices. In order to mimic the structural basis and coherent excitation energy funnelling in natural LHCs, it is viable to create biomimetic material hybrids with an optimal number of minor amorphous chromophoric units that peripheralize and couple effectively with a larger yet lower-energy core as an acceptor; in this way, the randomly-oriented peripheral chromophores can facilitate the maximization of photon absorption for all light polarization states[6] while an appropriate energy landscape in an engineered donor-acceptor molecular network gives rise to optimal excitation energy transfer mediated by ultralow-loss three-dimensional (3-D) FRET.[7,8] This demands fine tuning of the physical size, molecular arrangement, and electronic coupling within and between the donor and acceptor constitutes.

In contrast to the chemical homology naturally selected in LHCs,[9] the use of chemically distinct species has been employed in a variety of organic semiconducting devices in order to imitate the excitation energy funnelling and biophotocatalysis in LHCs. However, engineering chemical structures itself typically causes non-negligible overpotentials and energy losses,[10-15] owing to the presence of intrinsic energy-gap mismatch and heterogeneous interfaces;[16-19] however, this could be mitigated or remedied by utilizing the tunability of physical structures of the same chemical species. To do so, developing techniques that allow us to engineer the physical structures of a high-performing polymeric semiconductor (with spatial repetition of an identical chemical structure along its backbone) into an efficient 'self-doped' biomimetic donor-acceptor system would be a promising endeavour in our quest for ultrahigh light-harvesting or energy transfer efficiencies. This calls for realization of spatial definition of chain conformation, interchain packing, and macroscopic domain texture in a cost-effective device-relevant layer of the same conjugated polymer.[20-25] In this regard, solution-processable conjugated copolymers can be best-in-class multi-chromophoric systems, especially when the effective $\pi$-conjugation is coherently interrupted due to twisted or bended backbones.[6,26] This site-dependent structural feature tends to permit a straight chain segment of a polymer backbone to act as either a chromophoric or a fluorophoric element, depending on energy levels that are sensitive to the degree of chain extension and interchain packing structure. Therefore, a properly engineered polymeric semiconductor system can resemble a favorable host-guest system imitating the almost lossless energy transfer and structural regulation of photosynthetic organisms. For example, self-doping a polymer layer by tailoring its intrachain conjugation, interchain packing and electronic/vibrational coupling, would facilitate selective activation and more efficient transport of the competing excited interchain and interchain states.[6] Within this framework, molecular organization by virtue of the inherent long-range orientational order present in a liquid crystal (LC) mesophase of light-emitting liquid crystalline conjugated polymers (LCCPs)[27,28] is highly desirable.

Here we report the fabrication and systematic tuning of nematic polydomain and monodomain films of thermotropic poly(9,9-dioctylfluorene-co-benzothiadiazole) (F8BT). Its order parameter ranges from 0 to ≈100%, and the relative weighting between its intrachain and interchain light emission are modulated through the long-range orientational ordering in the nematic mesophase. A highly efficient host-guest system is thus generated from this novel self-doping method, whereby the excitation energy funnelling and photoluminescence quantum efficiency (PLQE) are remarkably enhanced. The relative fraction of (disordered) amorphous chromophoric host in the self-doped F8BT nematic polydomain film is also engineered to boost PLQE up to >70%, relative to a PLQE of ~30% in the spin-coated non-LC analogue. The improved light-harvesting and energy-conversion efficiencies in these bi-phase nematic polydomain films arise potentially from the synergy between the promotion of intrachain emission in F8BT nanocrystals as acceptor and the deactivation of energetic traps existing in a major faction of the amorphous matrix of F8BT chains, mediated by a biomimetic FRET process from the amorphous chromophoric host to a small fraction (<5%) of randomly oriented F8BT nanocrystals as the guest inclusion. However, further increases in overall crystallinity to 60% ~ 70% and order parameter to $S \approx 1$, as a result of high-quality photoalignment of chain orientation in the high-temperature nematic phase of F8BT and then freezing the structural order into a nematic monodomain glass film by a quenching step, is found to produce only an intermediate PLQE of ~50% as a result of notably elongated interchain packing coherence and emergence of an additionally allowed emissive interchain species whose PLQE is projected as <15%. The benefits and huge potential of engineering the relative ratio of F8BT nanocrystal acceptor in LC-organized F8BT polydomain glass films are showcased via tailoring the spatially averaged size of nematic polydomains by control over the film thickness. Thus, this study exemplifies the feasibility of utilizing a self-doped LCCP system to complement previous investigations of the structure-property relationship in non-oriented F8BT and related copolymers, whilst also reporting progress towards biomimicry of the unique structural basis and energy transfer properties of LHCs.

## 2. Results and Discussion

### 2.1. Materials and Film Preparation

The pristine LCCP samples were prepared by spin coating F8BT on a pre-cleaned Spectrosil substrate to form films with thickness ranging from 40 nm to 480 nm. Prior to the deposition of an F8BT layer, a continuous (discontinuous) ultrathin layer of our photoalignment material, namely hydrophilic sulphonic azo-benzene dye (SD1), was pre-coated directly onto substrate using a 0.5 mg/ml (0.1 mg/ml) solution and then aligned using linearly polarised ultraviolet (UV) illumination at $\lambda$ = 365 nm from an LED. The polarised UV illumination orients the long molecular axis of SD1 molecules in the plane of the substrate along a direction that is orthogonal to the polarization of the UV light source.[27-29] Subsequently, thermotropic LC-alignment or photoalignment of polymer chain orientation in an F8BT film or an SD1/F8BT bilayer was implemented by heating F8BT to its nematic mesophase before rapidly quenching to room temperature into a nematic glass (solid) film. This quenching step prevents polymer crystallization and thereby locks-in the self-assembled nematic director configuration, that is, a polydomain LC texture in a heated single layer of F8BT that has been directly coated on a substrate with no alignment layer [termed as *nonaligned* (NA) F8BT nematic glass film; labelled as *Film I*], and a monodomain texture for an F8BT film coated on a UV-aligned SD1 photoalignment layer [termed as *fully aligned* (FA) F8BT nematic films; labelled as *Film II*]. These highly oriented nematic monodomains are made possible by the long-range orientational ordering in the nematic phase via molecular interaction between the UV-aligned continuous SD1 layer and F8BT polymer chains at the interface, which is then communicated through the bulk of the LCCP layer by chain-chain interactions and $\pi$-$\pi$ stacking.[27] In addition, *partially-aligned* (PA) F8BT nematic glass films (labelled as *Film III*) were fabricated by quenching a heated overlying F8BT film that has been coated on a UV-aligned discontinuous SD1 layer resulting from spin-coating on a quartz substrate with a low concentration (herein 0.1 mg/ml) SD1 solution. Finally, the extensively studied *spin-coated* (SC) F8BT films were fabricated on a quartz substrate but without going through the nematic phase for use as control samples (denoted as *Film 0*) for microstructural and photophysical characterization. All F8BT films studied in this work are 160 nm in thickness unless stated otherwise. See Fig. 1 for the chemical structures of F8BT and SD1 and Fig. S2 for a schematic illustration of the four types of comparative F8BT films.

### 2.2. Microstructural Characterization and Chain-packing Model

Polarized optical micrographs (POMs) were recorded to visualize the LC texture formed in the F8BT films when each sample was placed between a crossed polarizer and analyzer pair. An F8BT nematic domain shows a dark (or bright) state when the local chain orientation (which defines the optic axis) is parallel (or at $45^{o}$) to the transmission axis of the polarizer or analyzer, whereas the spin-coated non-LC *Film 0* appears completely dark because of no birefringence effect. A typical Schlieren polydomain texture (*i* and *ii* in Fig. 1a) is observed in the mesophase self-assembled F8BT glass film (*Film I*). The direction of fluorophoric F8BT backbones in each nematic micro-domain is observed to be well-oriented by the LC long-range orientational ordering, evidenced by the highly polarised fluorescence in Fig. 1b; the local chain orientations of luminescent F8BT among different nematic micro-domains, however, are still randomly distributed, as confirmed by the varying fluorescence intensity across the polydomain texture.

High-quality chain alignment in a large-area-extended nematic monodomain is highlighted in terms of the uniform bright-state and dark-state POMs (*v* and *vi* in Fig. 1a) of the fully-aligned F8BT nematic glass film (*Film II*).

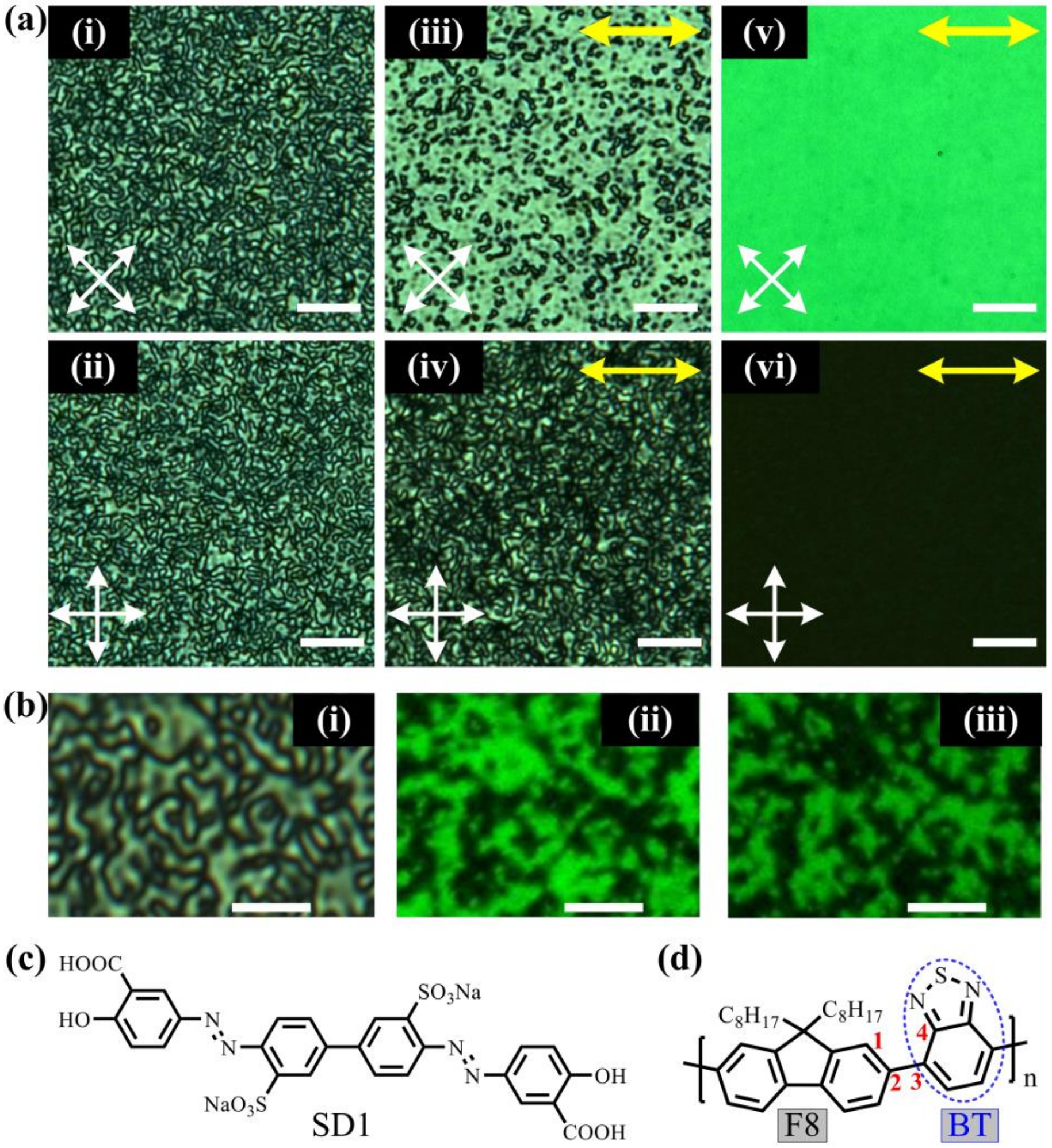

**Fig. 1** Observation of distinct LC textures in three comparative F8BT nematic glass films. (a) Polarizing optical micrographs (POMs) demonstrating a polydomain LC texture (*i* - *iv*) in both the nonaligned (*Film I*) (*i* and *ii*) and partially-aligned (*Film III*) (*iii* and *iv*) nematic films, and a monodomain texture in the fully-aligned nematic film (*Film II*) (*v* and *vi*), viewed between crossed polarizers (labelled by the two crossed white arrows). The photoaligned chain-orientation direction is indicated by the double-headed yellow arrows in (*iii*) - (*vi*), and all scale bars are 20 µm. (b) A POM image at a higher magnification showing a typical Schlieren texture in a nonaligned nematic *Film I* (*i*), and Corresponding polarized confocal fluorescence images (*ii* and *iii*) recorded when F8BT light emission was collected with the analyzer aligned to two orthogonal (i.e., horizontal and vertical) polarisation directions. The fluorescence was photoexcited with randomly polarized light at 450 nm; the scale bars in (b) are 5 µm. (c - d) Chemical structures of SD1 (c) and F8BT (d). The blue ellipse in (d) encloses the BT moiety and the four carbons labelled by the red numbers resemble a dihedral (i.e., intermolecular torsion) angle linking the F8 and BT units through a single carbon-carbon bond around which the F8 and BT moieties have a certain degree of freedom to rotate relative to each other.

Grazing-incidence wide-angle X-ray scattering (GIWAXS) measurement (Fig. 2a-d and Fig. S3) was carried out to clarify the microstructure and interchain packing in the F8BT films resulting from the different fabrication processes. The 2-D GIWAXS patterns for both the nonaligned polydomain *Film I* and the fully-aligned monodomain *Film II* are shown in Fig. 2a-c**,** all of which were presented after subtracting the GIWAXS data of the spin-coated reference *Film 0* as the background in order to uncover the effect of LC-alignment and photoalignment on long-range (*i.e.*, small $q$) structural ordering. Overall, the GIWAXS patterns show a big difference in crystallinity level, interchain packing, and structural anisotropy in different F8BT films. For the spin-coated non-LC *Film 0*, we see no clear indication of long-range ordering, evidenced by the absence of a clear characteristic diffraction peak. The nonaligned nematic polydomain *Film I* exhibits a faint smearing ring (see Fig. 2a and the insert in Fig. 2d), which can be taken as a result of randomly distributed chain orientations in a small proportion of localized crystalline regions; this also points to the inclusion of ordered "polymer nanocrystals" in an amorphous phase matrix of disordered F8BT chains.[30]

The noticeably higher diffraction signals and asymmetric GIWAXS patterns from the fully-aligned nematic monodomain *Film II* strongly suggest a substantial increase in the degree of polymer crystallinity and the structural anisotropy. Fig. 2b-d shows a rather strong diffraction ring recorded with a parallel X-ray incident configuration, alongside two different diffraction arcs from a perpendicular measurement configuration - a smeared (100) arc and a sharper (010) arc. The appearance of the short (100) arc suggests formation of highly oriented backbones in the extended nematic monodomain and that the relative rotation of F8 and BT moieties is restricted with both chemical units being aligned largely parallel to the plane of substrate. The sharper (010) arc at $q_{xy}$ = 1.055 Å$^{-1}$ gives an interchain repeat distance of 6.0 Å, which is a rather small periodicity when taking into consideration the length of the long alkyl side chains; this interchain repeat distance also corresponds to a condense interchain packing pattern with prevalent BT-to-BT close contacts between adjacent chains.

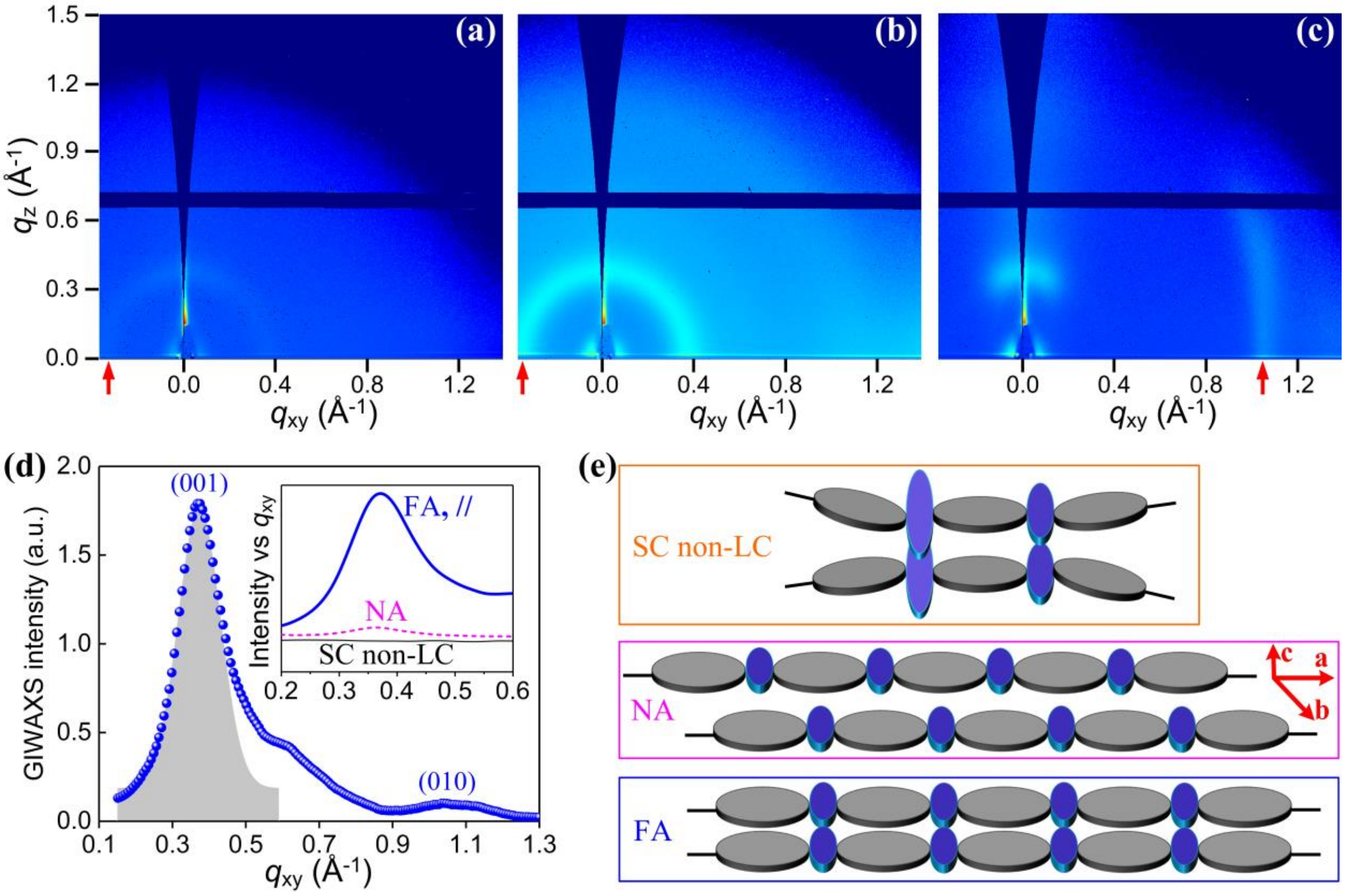

**Fig. 2** (a and c) 2-D GIWAXS images of (a) a nonaligned F8BT nematic polydomain *Film I*, and (b - c) a fully-aligned F8BT nematic monodomain *Film II*, recorded with the incident X-ray beam propagating in the SD1-aligned film along a direction parallel (b) or perpendicular (c) to the chain alignment direction. The vertical red arrows on the bottom of (a) - (c) mark the $q_{xy}$ position of the characteristic scattering peaks. (d) In-plane linecut GIWAXS profile (blue dot) of the fully-aligned *Film II* after subtracting a second order baseline, recorded with the parallel (//) configuration. The shaded region under the as-shown Gaussian peak fitting curve in (d) represents the area of the (001) peak used to estimate the degree of overall crystallinity; the inset plots the raw scattering profile (offset for clarity) of the fully-aligned (FA from the // configuration) and nonaligned (NA) F8BT nematic glass films and of a spin-coated (SC) F8BT film (*Film 0*). (e) Schematic illustration of the proposed model for polymer chain-conformation and interchain packing structures in the three types of F8BT films. In (e), the grey and blue ellipses denotes F8 and BT moiety along an F8BT backbone, where the side-chains and backbone kinks are omitted for clarity; *a*, *b,* and *c* axes in the middle panel are shown along the backbone, interchain, and out-of-plane direction, respectively.

The *q* values of the characteristic diffraction peaks and extracted *d*-spacing and coherence lengths for the GIWAXS results are listed in Table S1. Altogether, these diffraction results substantiate the feasibility of utilizing mesophase self-assembly to induce polymer nanocrystal dispersions in the nonaligned F8BT nematic *Film I*, where an extended chain conformation with a slightly longer (by 3%) monomer length along the backbone direction and larger but 3-D isotropic coherence lengths (141.7 - 149.6 Å) is locked-in. The highly oriented nematic monodomain exhibits highly anisotropic coherence lengths, i.e., 120.8 Å, 205.1 Å, and 101.0 Å along the backbone, interchain, and film thickness directions, respectively. The degree of polymer crystallinity over the representative photoaligned F8BT nematic monodomain was estimated to be 60% ~ 73%, quantified in terms of the area ratio of the (001) and (010) peaks in the total diffraction intensity when we took the shaded region under the Gaussian peak fitting curve in Fig. 1d as the area of the (001) diffraction peak. The fraction of F8BT nanocrystals in the nonaligned polydomain *Film I* had an estimate of 4% - 5% calculated from on the relative intensity of the faint (100) ring in the insert of Fig. 3d.

The spontaneous mesophase structural self-reorganization in the nonaligned polydomain *Film I* involves the stabilization of a minor fraction of the highly-ordered polymer nano-crystalline nuclei (which have a lower energy bandgap) within a pool of disordered nematic F8BT chains that have a chain length distribution defined by polydispersity. We expect these highly localized crystallization events to initialize through LC self-assembly and associated chain-extension of the most rigid (or straight) chain segments on the longest F8BT backbones, which would continue to selectively incorporate the next most extended chains in the mesophase into the already crystallized cores.[30] For the fully- and partially-aligned nematic F8BT films, a similar LC-alignment process may proceed via a gradual expansion of the polymer nanocrystals at the expense of domain boundaries and re-orientation of the newly incorporated F8BT chains to a director direction defined by the UV-aligned SD1 commanding layer.

On the basis of all of our GIWAXS and POM results, an F8BT packing model has been proposed for different F8BT films (schematically illustrated in Fig. 2e) to explain their different structural features and photophysical properties (*vide infra*). For the spin-coated non-LC film (*Film 0*), BT moieties have relatively random orientations but a high degree of intermolecular torsion with respect to the F8 unit (inherited from the significant structural disorders in an F8BT solution) and remain close to each other in neighboring chains to minimize steric hindrance. This kind of highly twisted molecular arrangement agrees with the dominance of amorphous phase and significant BT-to-BT close contacts identified in *Film 0*. On the contrary, LC long-range chain ordering induces polymer nanocrystals in the nonaligned nematic polydomain *Film I* through the re-structuring of F8BT chains into a more energetically favorable configuration, via adopting more planar chain-conformations and dramatic depression of intermolecular distortions; in this case, BT moieties in one F8BT chain would prefer to occupy positions adjacent to the F8 location of a neighboring chain (thus recoiling the so-called "alternating structure") due to the repulsive force between BT-to-BT inter-chain dipoles.[31] Analogously, the fully-aligned monodomain *Film II* can favour a more extended and rigid chain-confirmation but, in this case, the highly oriented F8BT chains have translated relative to each other by half a repeat unit along the backbone direction; this aids the promotion of a higher number density of close interchain BT-to-BT contacts, particularly those under the templating effect imposed by the UV-oriented (and probably cross-linked) SD1 molecules.

## 2.3. Photophysical Characterization

The inherent mesophase long-range orientational ordering of light-emitting LCCPs is able to create a self-doped host-guest system, in which the disordered amorphous (high-energy) host acts as the chromophoric donor while the self-assembled polymer nanocrystals become the fluorophoric or acceptor constituent.[26,27] While the amorphous host and polymer nanocrystals both contribute to the dielectric and optical absorption behaviours of a bi-phase film, the self-doped polymer nanocrystals would dictate the emission properties by facilitating nonradiative host-to-guest FRET and deactivating possible traps in the major amorphous host. As a result, tailoring the effective π-conjugation, the spatial distribution, and relative weight of the host component would offer an intriguing means to fine-tune the optical, electrical and photophysical properties of light-emitting LCCPs, which seems feasible for mimicking the ability of light-harvesting antennas to regulate the number ratio of active minor- and larger-sized LHCs in response to fluctuating sunlight.[1,5] In the case of LCCP F8BT, the lowest unoccupied molecular orbital (LUMO) localizes on the BT moieties, and the highest occupied molecular orbital (HOMO) delocalizes over the entire backbone.[31-34] Therefore, the absorption, migration and decay of absorbed photon energy in the LC-assembled and photoaligned F8BT glass films are fundamentally governed by the intrinsic competition among the structurally allowed intrachain and interchain excited states.

Polarized optical absorption spectra are collected to elaborate the electronic states and optical transitions in the comparative F8BT films. The absorption spectrum from the nonaligned F8BT nematic polydomain *Film I* (Fig. 3a) demonstrates a broadening of the lowest-lying absorption band towards longer wavelengths, indicating the emergence of a new lower-energy absorption species corresponding to F8BT nanocrystals therein. The peak wavelength of this first-allowed optical transition band was blue-shifted from 498 nm in the nonaligned polydomain *Film I* to 486 nm in the partially-aligned nematic polydomain *Film III*. This is a signature of continuing incorporation of higher-energy chains during the growth of polymer nanocrystals; the peak wavelength further reduces to 465 nm in the non-LC *Film 0* and to 460 nm in the fully-aligned F8BT nematic monodomain *Film II* as a result of the spatial average effect of a range of chain lengths. The absorption strength of the extended nematic monodomain is seen to be augmented relative to both the nonaligned polydomain *Film I* and the non-LC *Film 0* of the same thickness, since the photoaligned backbones could remain largely parallel to the substrate plane and the chains in F8BT nanocrystals while amorphous matrices are randomly oriented, resulting in the attenuation of oscillation strength when measurements are taken along the thickness of the film. Domain boundary regions separating nematic polydomains are expected to affect attenuated absorption strength similarly. The long-wavelength spectral tail of the nonaligned polydomain *Film I* is ascribed to light scattering due to the structural heterogeneities associated with the randomly-oriented nanocrystals.

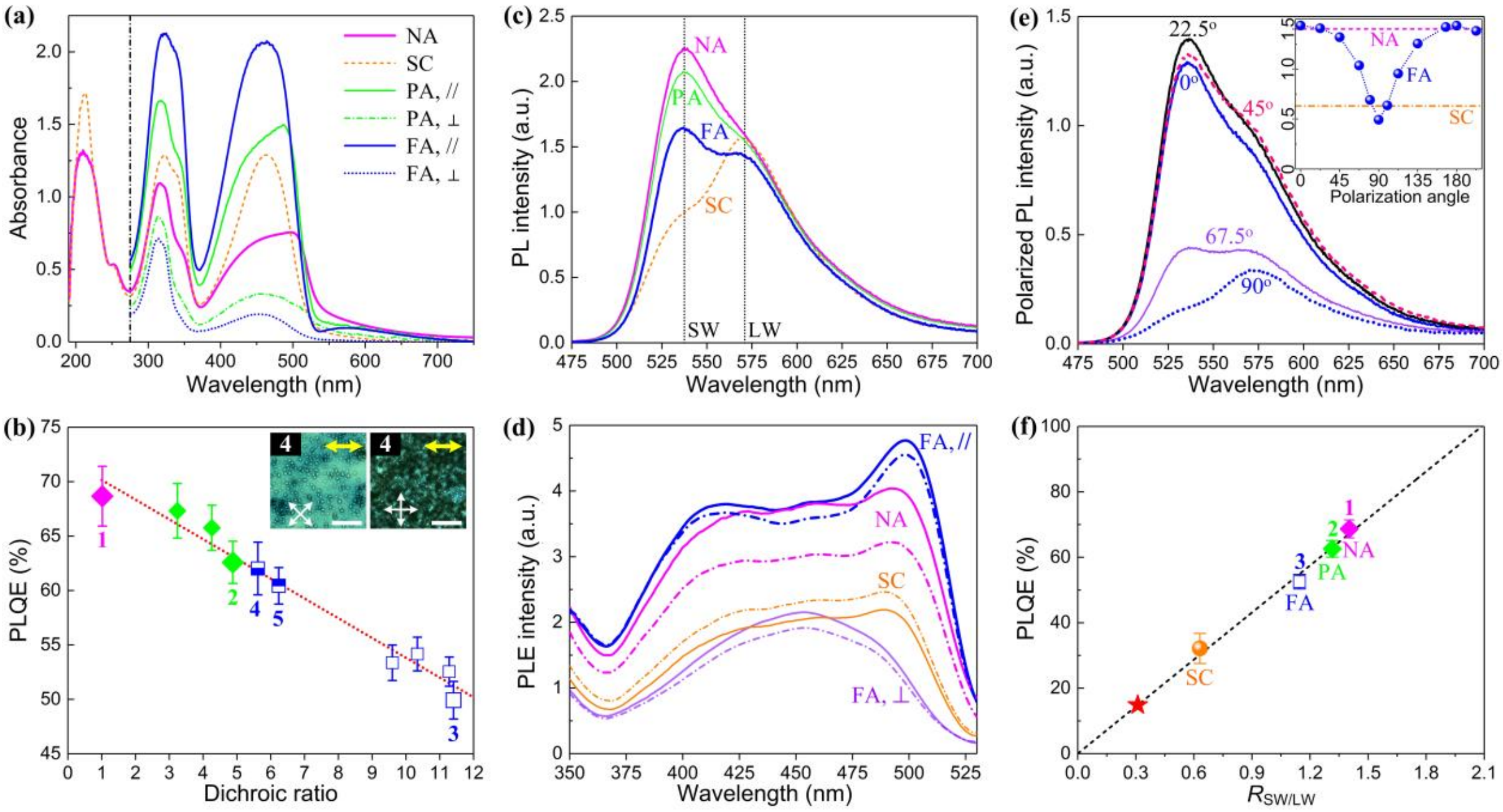

**Fig. 3** Photophysical characterization of the spin-coated (SC) non-LC F8BT film and of the nonaligned (NA), partially-aligned (PA) and fully-aligned (FA) F8BT nematic glass films. All parallel (//) and perpendicular (⊥) polarised spectra in the PA and FA films were recorded by aligning the excitation polarization parallel and perpendicular to the chain alignment direction, respectively. (a) Optical absorption spectra of the four comparative F8BT films; the polarised spectra were cut off at 275 nm below which the transmission of the polariser drops rapidly, whereas the non-polarised spectra for both the SC reference film and the NA nematic film were extended to 190 nm. (b) Plot of PLQE in different nematic glass films of F8BT vs optical dichroic ratio determined at 460 nm. The data presented by the four blue opened boxes comes from the fully-aligned nematic monodomain films, and the fully-filled diamond and half-filled box data comes from the nematic polydomain films of either NA (#1) and PA (#2) type, or a fully-aligned but thinner F8BT film [e.g., the sample #4: 40 nm thick, and sample #5: 80 nm thick]. The inserts in (b) display the bright-state (left) and dark-state (right) POMs of the sample

#4, with scale bars being 20 µm. (c) Non-polarized PL spectra and (d) Polarized-excited and nonpolarised- collected PL excitation (PLE) spectra of various F8BT films. The two vertical dotted lines in (c) mark the identified shorter-wavelength (SW) and longer-wavelength (LW) PL peaks at ~536 nm and 570 nm, which are assigned to the center of two vibronic optical transition bands. In (d), all solid curves were recorded for emission at 536 nm while the dotted curves for emission at 570 nm. (e) Polarized PL spectra of the fully-aligned nematic monodomain *Film II,* probed under randomly polarised excitation light source and when emission was collected with polarization aligned at the as-labelled polarization angles ($\theta$) relative to the chain alignment direction. The insert in (f) plots the PL intensity ratio of the monodomain *Film II* at the SW (536 nm) to LW (570 nm) vibronic peak, $R_{SW/LW}$, as a function of the emission collection polarization angle $\theta$, alongside the $R_{\mathrm{SW/LW}}$ of the nonaligned polydomain *Film I* (pink dashed line) and the spin-coated *Film 0* (orange dash-dotted line). (f) PLQE vs $R_{SW/LW}$ for the four comparative F8BT films, where the red star denotes an extrapolated PLQE for so-called "pure amorphous-phase F8BT". The dotted and dashed line in (b) and (f) is a guide for a linear and a proportional change trend, respectively. All PLQE and PL spectra were measured using the same randomly polarised light source at 450 nm.

The polarized absorption spectra of the fully-aligned *Film II* emphasizes high quality chain-orientation and uniform monodomain texture; this manifests in the form of an ultrahigh optical dichroic ratio [i.e., the absorbance ratio between parallel ($A_{//}$) and perpendicular ($A_{\perp}$) polarized spectra to the chain alignment direction] of 11.4 at the peak wavelength of the first absorption band and the equivalent order parameter of $S \approx 0.98$.[27] The combined effects of highly oriented backbones in the plane of the substrate and the elimination of domain boundaries in the fully-aligned monodomain *Film II* are essential for creating such an exceptional level of structural ordering and anisotropy in photophysical properties. The texture uniformity in *Film II* is also evident from the long wavelength interference fringes seen in its $A_{//}$ spectrum.

Overall, the absorption spectra in Fig. 3a are consistent with the identified microstructures and proposed F8BT packing structures. The absorption peak intensity ratio between the first (375 - 500 nm wavelength) and second (275-375 nm) optical transition band increases from ~2/3 in the nonaligned nematic polydomain *Film I* to levels close to 1 in both the spin-coated non-LC *Film 0* and the fully-aligned nematic monodomain *Film II.* This observation infers that F8BT chains in both *Film 0* and *Film II* are largely restricted in the plane of substrate, whereas F8BT chains in the polymer nanocrystals and amorphous matrix in *Film I* are randomly oriented in the 3-D space. Observations of an increase in the first and second transition peak intensity ratio with chain alignment quality would help to rationalize the following points: (*i*) spatial excitation energy transfer and concomitant first optical transition band associated with a charge-transfer character involve BT units through intrachain π-conjugation and/or BT-to-BT interchain coupling and (*ii*) both the intrachain and interchain interactions are potentially enhanced by the high-quality polymer photoalignment and BT-to-BT interchain alignment. In addition, the peak strength of the third transition band (centered around 200 nm, originating from the optical oscillation of benzene rings[31,34]) in the spin-coated *Film 0* is determined from Fig 3a to be 1.3 times that of the nonaligned nematic glass film, arising from a random three-dimensional (3D) distribution of F8BT chain orientations in the polymer nanocrystals and amorphous matrix. The second absorption band consisting of two transition constituents centered at 315 nm and 340 nm, manifests distinct vibronic structures in the four types of F8BT films, in agreement with their different π-π stacking of F8BT chains.[34] In particular, the lowest areal fraction of the 340 nm vibronic sideband in the second absorption band can be related with an alternating packing structure as proposed for the nonaligned F8BT nematic *Film I*, where fewer BT-to-BT close contacts and a larger extent of interchain misalignment are present; the disordered chain-conformation and interchain packing in the non-LC *Film 0* has a moderate relative weight of the said 340 nm vibronic sideband due to significantly more BT-to-BT close contacts between adjacent chains and enhancement in interchain interactions. Among all measured F8BT films, the fully-aligned monodomain *Film II* exhibits the highest fraction of 340 nm vibronic sidebands in the $A_{//}$ spectrum and also the lowest relative weight in the $A_{\perp}$ spectrum, which are in accordance with improved crystallinity and interchain coherent length.

In Fig. 3b, we illustrate an unusual linear dependence of the PLQE in the F8BT nematic glass films on the chain-alignment quality (parametrised as the optical dichroic ratio here). A maximum enhanced PLQE of 68.8% ± 2.6% is shown for the nonaligned nematic polydomain *Film I* in comparison with PLQE of 31.7% ± 4.7% in the spin-coated non-LC *Film 0*. While the PLQE values for the spin-coated layers of non-LC F8BT remain consistent with previously reports,[31,35] it is noteworthy that the difference between a PLQE of 71.6% ± 3.5% determined from F8BT solution and our maximized PLQE in the nonaligned polydomain *Film I* is within the measurement error. The nematic polydomains in either the nonaligned *Film I* or the partially-aligned *Film III* display a PLQE remarkably larger than that of the fully-aligned monodomain *Film II* (approximately 50%). This linear PLQE change trend is further supported by the non-polarised PL spectra shown in Fig. 3c and polarised PL excitation (PLE) spectra in Fig. 3d, in the form of a good agreement between the PL/PLE intensities and PLQE value. Although all of the measured F8BT films show a well-resolved vibronic PL peak at ~536 nm and another longer-wavelength vibronic optical transition band centered at ~570 nm, the relative PL peak intensity ratio of the short-wavelength to long-wavelength vibronic PL peak, $R_{SW/LW}$, differs in different films (see also Fig. S4a for their normalized PL spectra): the non-polarized PL spectra from both the nonaligned polydomain *Film I* and the partially-aligned polydomain *Film III* are dominated by the first allowed (0-0) vibronic transition band, while the fully-aligned monodomain *Film II* shows a relatively low $R_{SW/LW}$ value which is still greater than one; on the other hand, the spin-coated *Film 0* exhibits the largest fraction of the longer-wavelength vibronic transition in the PL spectrum.

These distinct $R_{SW/LW}$ results and PL peak intensities among all F8BT films are in line with their PLE spectra acquired for emission at 536 nm and 570 nm (Fig. 3d). With the exception of *Film 0*, all F8BT films give a stronger PLE intensity for the 0-0 vibronic PL peak than emission at 570 nm. Considering that *Film 0* has a nearly amorphous feature and no long-range exciton migration pathways, its highly twisted chain conformations would favour relatively lower-energy optical transitions. Entering the LC mesophase in all the LC-assembled F8BT glass films, on the other hand, improves PLQE by promoting the photo-excitation of higher-energy (0-0) vibronic emission because of enhanced intra/inter-chain coherence lengths and mitigated chain entanglements. For instance, an ensuing increase in effective π-conjugation length delocalizes excitonic wavefunction and allows access to extra emissive intrachain and interchain species of collective oriented F8BT chains, evidenced by the appearance of a new lower-energy PLE sub-band peaking at a longer wavelength that we have shown for the nonaligned and fully-aligned nematic glass films (see Fig. 3d).

In contrast to *Film 0* and nonaligned nematic polydomain *Film I* whose PLE spectra are almost independent on the excitation polarisation, the fully-aligned nematic monodomain *Film II* shows strongly polarisation-dependent PLE properties in correspondence to the anisotropic structural order and improved intra/inter-chain coherence lengths. Under parallel polarised photo-excitation, the PLE spectra in the extended nematic monodomains recorded for emission at 536 nm and 570 nm, give stronger emission across the narrow longer-wavelength sub-band and also a PLE plateau extended to shorter wavelengths relative to the PLE spectra of the nonaligned nematic polydomain and non-LC films. The clearly extended PLE spectrum points out the feasibility of high-quality and long-range polymer alignment for mitigating energetic disorder and activating band-edge and intra-band states.[36] A decrease in the range of ordered interchain packing such as in F8BT nanocrystals allows to promote only additional band-edge states, probably due to increasing energetic disorder along the interchain direction prior to the occurrence of light emission. When pumping with perpendicular polarised light, the two PLE spectra of the fully-aligned *Film II* became relatively featureless, narrowed in wavelength range and lower in intensity because the reduced optical absorption may be insufficient to activate the sub-band and band-edge states. Spatial averaging of the parallel and perpendicular polarised PLE spectra in *Film II* lowers the total PLE and PLQE to a level less than these of the nonaligned nematic polydomain *Film I*.

The anisotropic PLE spectra and intermediate PLQE of the fully-aligned F8BT nematic monodomain *Film II* commensurate with the polarised PL spectra shown in Fig. 3e. The PL intensity of the 0-0 vibronic transition band is augmented when emission was collected with a polarisation direction at a smaller angle (i.e. $\theta_{Em}$ = 0° - 45°) to the chain alignment direction, and is maximized at $\theta_{Em}$ = 22.5°. There also exists a gradual shift in the $\theta_{Em}$-dependence of relative PL peak intensity between two vibronic PL bands centered at 536 nm and 570 nm (see the insert in Fig. 3e), with the lower-lying vibronic transition band maximizing its fraction at $\theta_{Em}$ = 90°. Also shown in Fig. 3e is the intensity of the 45°-polarised PL spectrum exceeding the 22.5°-polarised PL spectrum at wavelengths of >553 nm; we attributed this phenomenon to a competition between the gradually reduced intensity of the linearly polarised intrachain emission as $\theta_{Em}$ increases from 22.5° to 90° and a simultaneous increase in intensity of an emissive excited species with an interchain character.[27] Accordingly, the different $\theta_{Em}$-dependence of the two orthogonal PL components is accompanied by a more complicated angular variation of $R_{\mathrm{SW/LW}}$ than a simple $\cos^2(\theta_{Em})$ relation that is expected for a single polarised emissive spices. We propose that the discrepancy of the fluorescent excited intrachain and interchain species can be correlated with possible prohibition of the 0-0 vibronic transition of interchain emission, in comparation with emissive intrachain species that is shown to allow 0-0, 0-1, and 0-2 vibronic transitions centered at around 536 nm, 553 nm and 570 nm, respectively. Altogether, our vibronic PL spectral features, along with the polarisation-dependent PLE properties in the fully-aligned monodomain *Film II*, encourage further assignment of *J*-type and *H*-type aggregation[25] to the emissive excited intrachain and interchain state, respectively. Note that *J*-aggregation can trigger superradiation while typical *H*-aggregates induce a far lower PLQE.[37]

Additional polarised PL spectra of the fully-aligned F8BT nematic monodomain film, detected using four combinations of excitation ($\theta_{Ext}$) and emission collection ($\theta_{Em}$) polarisation angle relative to F8BT chain alignment direction, are provided in Figs. S4b and S5 to underline ultrahigh PL polarisation anisotropies: a total PL intensity ratio of 28.6 and a PL peak intensity ratio of 40.4 at 536 nm between PL spectrum detected at $\theta_{Em}$ = 0°, $\theta_{Ext}$ = 0° and $\theta_{Em}$ = 90°, $\theta_{Ext}$ = 90°. These exceptional PL polarisation anisotropies showcase unique advantages of making use of the polarization degree of freedom for selective photo-excitation and emission in highly oriented LCCPs. Also highlighted in Fig. S4b is a nearly identical vibronic spectral lineshape between the 22.5° (and 0°)-polarised PL spectra of *Film II* and the nonpolarised PL spectrum of *Film I*, implying a common origin of PL emission in the two films, that is, the radiative decay of emissive excited intrachain state in the fully-aligned nematic monodomain *Film II* and the LC-assembled polymer nanocrystals in the non-aligned polydomain *Film I*. These findings also support non-radiative FRET funnelling from the amorphous host to polymer nanocrystals in the nematic polydomains and therefore suppressed emission from the major amorphous matrix.

By taking the $R_{SW/LW}$ ratio at the 0-0 (536 nm) and 0-2 (570 nm) vibronic peak as an estimation of the relative weighting of intrachain emission in a non-polarised PL spectrum, we can underline a proportional relationship between the PLQE of all comparable F8BT films and the corresponding $R_{SW/LW}$ in in Fig. 3f. For the spin-coated non-LC *Film 0*, highly twisted chain-conformations and highly localized BT-to-BT interchain interactions explain the observed dominance of the 0-2 vibronic transition band (thus, the smallest $R_{SW/LW}$) and the lowest PLQE. The highest PLQE and greatest fraction of intrachain emission obtained for the nonaligned nematic polydomain *Film I* can be ascribed to establishment of a self-doped host-guest system, which combines the benefits of optimal absorption of non-polarised photon energy, highly efficient host-to-guest FRET, dominant intrachain emission, and deactivating traps and energy losses associated with a major fraction of disordered F8BT chains. Even though the monodomain induces an additional emissive interchain species with an *H*-aggregation character, a spatial averaging of this weakly emissive interchain species and efficient intrachain emission leads to only intermediate $R_{SW/LW}$ and PLQE under nonpolarised excitation. Nevertheless, when the intrachain emission was spectated by detecting PL signal with detected at $\theta_{Em}$ = 0, the $R_{SW/LW}$ ratio is found to be up to 2.11~2.15 in the most ordered fully-aligned LC monodomain thin films (**Fig. S6**), pointing to a near-unity PLQE for the intrachain emission.

Taking all of the structural and photophysical spectral information into account, we rationalize *i*) that a combination of 3-D host-to-guest FRET funnelling and *J*-aggregated intrachain emission in F8BT nanocrystals governs the desirable PL properties in the self-doped F8BT nematic polydomain glass film, *ii*) that high-quality chain orientation in the extended nematic monodomains brings about high anisotropy in charge-carrier transport and polarised absorption/emission that are inaccessible in the non-LC and non-aligned nematic polydomain analogues, whilst also promoting a new emissive excited interchain species that leads to an orthogonal polarised 0-1 vibronic PL emission band and reduction in the spatially-averaged PLQE because of the *H*-aggregation character and non-separativie spectral feature from intrachain emission, and *iii*) that the structural and energetic disorder in the spin-coated non-LC *Film 0* and the major amorphous matrix of the nonaligned nematic polydomain *Film I* plays distinct roles in the overall PL performance: the amorphous F8BT chains still dominate the optical and photophysical properties in *Film 0*, but the LC alignment upgrades the PLQE and effective defect tolerance of the nematic polydomains. Moreover, disordered F8BT chains in the non-LC *Film 0* result in a wider 0-2 vibronic PL band [e.g., full width at half maximum (*FWHM*) = 71 nm] than for the remained non-aligned nematic-phase F8BT chains (*FWHM* = 46 nm) in the fully-aligned monodomain *Film II* (c.f. Fig. S7), suggesting that the LC-alignment mitigates energetic disorder in the donor matrix. Based on the $R_{SW/LW}$ data obtained by fitting the 0-2 vibronic PL transition band in the non-aligned nematic F8BT chains (Fig. S7a), we extrapolate the PLQE of "pure amorphous" nematic F8BT to be ≈15% (the red star in Fig. 3f). This presents a lower PLQE value by a factor of ~5 and ~7 compared respectively with F8BT intrachain emission in LC multidomains and monodomains, hence agreeing with the *H*-aggregation character in the amorphous nematic F8BT.

### 2.4 Polarized Micro-PL Spatial Spectral Mapping

High-resolution spatial definition of the chain-orientation, microstructural anisotropy and concomitant energy landscape can be used to tailor the photophysical properties in the different types of F8BT films. The specific extent of structural inhomogeneity and energetic disorder, on the other hand, is extremely sensitive to the probing location and length scale of the relevant photophysical measurements. In order to elucidate the effect of high-quality chain alignment and self-assembled LC textures on the localised emission properties (e.g., polarised PL spectra and vibronic spectral lineshape, PL intensity anisotropy), systematic polarized micro-PL (μ-PL) measurements (c.f. Fig. S1 for the experimental setup) were carried out on an SD1-aligned, spatially-patterned 3 μm-thick F8BT line against the nonaligned polydomain background in the same 160 nm-thick F8BT nematic glass film sample. Spatial patterning of chain-orientation in the sample, as shown in the bright-state POM in Fig. 4a and Fig. S8, was created using UV-alignment with a photomask to illuminate a continuous SD1 layer before accessing the high-temperature nematic phase to selective alignment of the overlying F8BT film (see Experimental Method and Fig. S8 for more details). It has been demonstrated that 2 - 4 μm feature sizes are readily generated in the photo-masked chain-orientation patterns against a nonaligned background in the correspondingly photoaligned F8BT nematic glass film.

All polarised μ-PL spectra were acquired using a 405 nm excitation source with fixed beam polarization parallel to the chain-orientation direction (i.e. $\theta_{Ext} = 0°$) in the UV-aligned line (along the *y*-axis in Fig. 4b-c). The emission was separated using an analyzer with polarization parallel ($PL_{//}$) and perpendicular ($PL_{\perp}$) to the transition dipole moment direction of the highly oriented F8BT chains in the patterned line region. By rotating the emission collection polarisation in the film plane during acquisition of the polarised μ-PL spectra within the aligned line, the direction of F8BT transition dipole moment (marked by the red dashed arrows in the insert of Fig. 4e), where μ-PL intensity from the aligned F8BT line is maximized under the fixed excitation, is found to align at $\theta_{Em} \approx 20^{o}$ relative to the chain-orientation direction in the aligned channel. This deviation angle is close to the experimental and computational results of $20^{o}$ - $22^{o}$.[38,39] A 3-D rendered map (bottom of Fig. 4b) and a 2-D colour-coded image (top of Fig. 4b) are included to help visualize the spatial variation of the integrated $PL_{\perp}$ spectral intensity in the aligned line and the nonaligned polydomain background (emission from the background was collected with the same excitation and analyzer polarization used to record the $PL_{\perp}$ spectra of the aligned line); the corresponding 3-D mapping of the integrated $PL_{//}$ intensity from the same region of the sample is provided in Fig. S9a. The nonaligned polydomain background displays an irregular variation in the domain size/shape and μ-PL intensity along both *x*-axis and *y*-axis, revealing a random distribution of the chain orientation across different nematic polydomains.

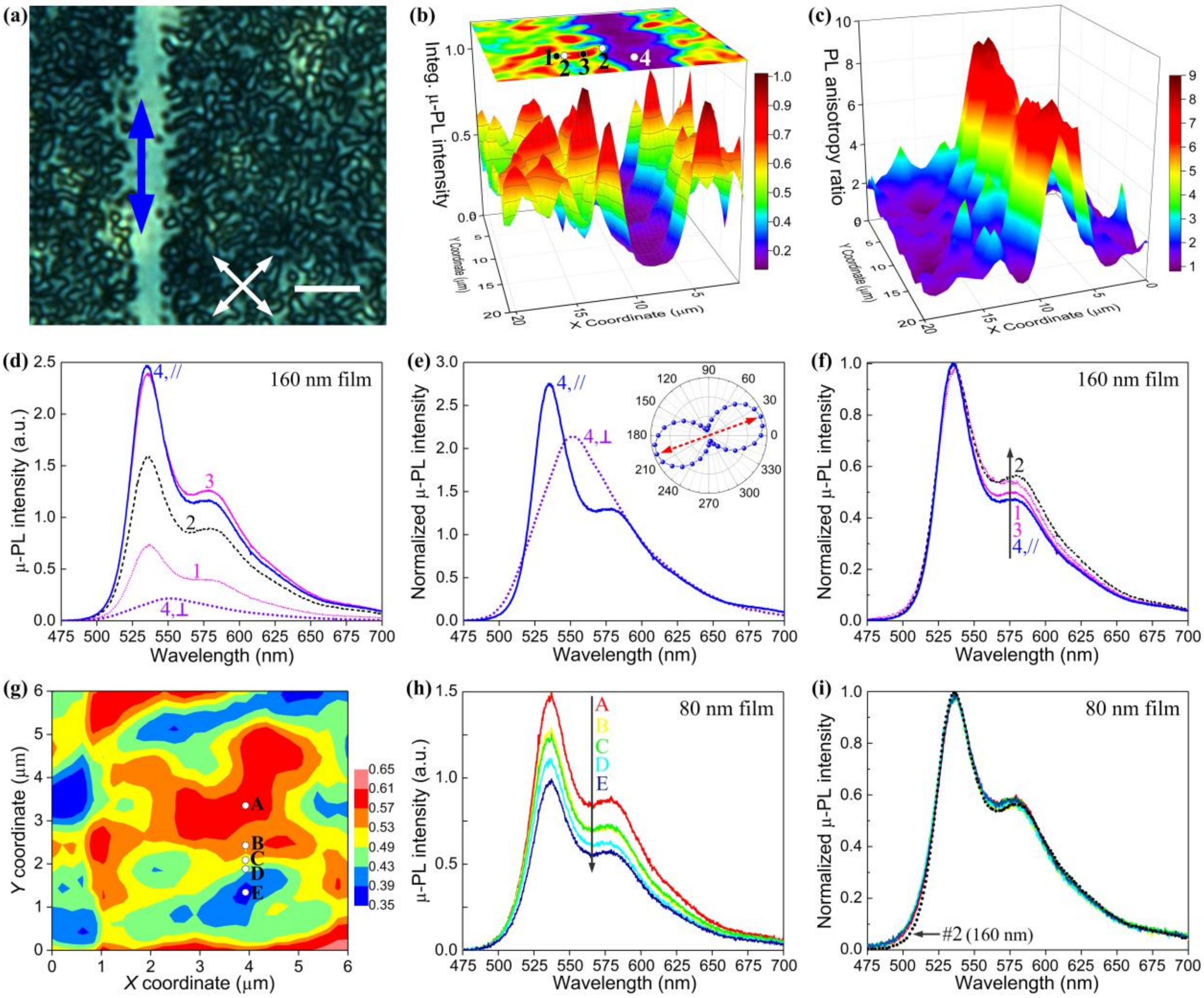

**Fig. 4** Polarized µ-PL spectral characterization of (a - f) a spatially patterned 160 nm-thick and (g - i) a nonaligned 80 nm-thick F8BT nematic glass film. For the sample region plotted in (a) - (f), UV-alignment of an SD1 layer has been patterned with a photomask consisting of an aligned channel against a non-aligned background; the resulting F8BT alignment direction in the channel region is indicated by the double-headed blue arrow in the polarizing optical micrograph (a) recorded from the same spatially-patterned F8BT film (horizontal scale bar: 6 µm). (b) 3-D colour-coded µ-PL map and a contour plot of the normalized total µ-PL intensity as a function of the in-plane location, and (c) Corresponding 3-D map of the spatial variation of the integrated PL anisotropy ratio. (d) Polarised µ-PL spectra and (e - f) Corresponding peak-normalized µ-PL spectra acquired at the different locations in the polydomain region [labelled as #1 - #3 in (b)] and also at the aligned line (#4). The µ-PL spectra in (e) were normalized by the 0-3 vibronic peak intensity and these in (f) by the 0-0 vibronic peak. For the scanning point #4, the perpendicular (⊥) and parallel (//) polarised µ-PL spectra were recorded with emission collection polarization perpendicular and parallel, respectively, to F8BT transition dipole moment. The inset in (e) is a polar plot of the normalized total µ-PL intensity from location point #4 vs the angle between the emission collection polarisation and the chain alignment direction. (g) A contour plot of the normalized total µ-PL intensity across the polydomains in a representative nonaligned 80 nm-thick F8BT nematic glass film. (h) Polarized µ-PL spectra and (i) Corresponding normalized PL spectra for points *A*, *B*, *C*, *D* and *E* labelled in (g). The normalized µ-PL spectrum from the domain boundary [the thick black dotted curve from location #2 in (b)] in the 160 nm-thick nonaligned background was plotted in (i) for comparison.

Fig. 4c further illustrates relatively small values of integrated PL anisotropy, i.e., = $\int PL_{//} / \int PL_{\perp}$, from the nonaligned nematic polydomain background compared to the much larger and more uniform anisotropies in the photoaligned F8BT channel. The highly oriented F8BT chains in the aligned line collectively induce large integrated anisotropies ~9 due to the combined benefits of LC alignment and high-quality photoalignment, especially those that involve a slight enhancement of the (intrachain) $PL_{//}$ intensity but a noticeable reduction in the interchain-emission-dominated $PL_{\perp}$ intensity.[40] Inspection of the spatial variation of PL anisotropy across the LC-assembled polydomain background (see also Fig. S9b) estimates the average domain size to be 3.5 - 5 µm by reflecting a 90$^{o}$ difference of chain orientation formed in nematic polydomain domains, which therefore tallies with the irregular polarised confocal fluorescence images shown in Fig. 1b. For comparison, the averaged domain size of ~2 µm across the POM images of the nonaligned polydomain *Film I* relies on a 45$^{o}$ difference in chain orientation.

In order to visualize the spatially resolved vibronic PL spectral structures in the in-plane patterned sample of nematic F8BT, Fig. 4d-f (and Fig. S8c-d) show the polarised µ-PL spectra recorded from different locations across the representative domains in the nonaligned polydomain background (location #1 - #3) and in the photoaligned monodomain channel (location #4), as labelled in Fig. 4b. All of these highly localised µ-PL spectra should be dictated by the highly oriented F8BT chains or nanocrystals, given that emission of the disordered F8BT chains in the nonaligned polydomain film is effectively suppressed by nonradiative donor-to-acceptor FRET funnelling. Altogether, our polarised µ-PL spectral results unravel similar spectral lineshapes in the $PL_{//}$ and $PL_{\perp}$ spectra collected from the three probe locations in the nonaligned polydomain background, except for the domain boundary which (location #2) exhibits the best resolved (0-2) vibronic PL peak at ~580 nm and the biggest PL peak intensity ratio between the 0-2 and 0-0 vibronic transition, $P_{(0\text{-}2)/(0\text{-}0)}$. $P_{(0\text{-}2)/(0\text{-}0)}$ and μ-PL intensity are generally correlated such that the higher the μ-PL intensity is from a recorded location, the lower the $P_{(0-2)/(0-0)}$ ratio in the μ-PL spectra is for the same scanning location. Among all polarised μ-PL spectra, the $PL_{//}$ spectrum from location #4 in the photoaligned channel manifests the highest PL intensity at the 0-0 vibronic peak but the lowest $P_{(0-2)/(0-0)}$. Also, these μ-PL spectral findings agree with the shown relationship between $P_{(0-2)/(0-0)}$ and PL intensity from the macroscopic PL spectra in Fig. 3c and e.

In contrast to the analogous μ-PL spectral lineshapes recorded from the three typical locations in the nonaligned polydomain background, a striking PL-spectral separation of two orthogonally polarised PL components originating from emissive excited intrachain and interchain species is presented in Fig. 4e for the photoaligned nematic monodomain line. Although the $PL_{//}$ spectra from location #4 can resemble the vibronically-structured μ-PL spectra of the nonaligned polydomain background (but with a slightly lower $P_{(0-2)/(0-0)}$ ratio that is equivalent to a greater Huang-Rhys parameter and more rigid molecular geometry of the aligned chains), the relatively featureless $PL_{\perp}$ spectrum from location #4 peaks at ~553 nm and has a much lower PL intensity than the $PL_{//}$ spectrum, having a PL intensity ratio of 15.4 at the 0-0 vibronic peak. When normalized by the 0-3 vibronic PL peak intensity, the $PL_{//}$ and $PL_{\perp}$ spectra from location #4 in Fig. 4e display distinct 0-0 and 0-1 vibronic PL peak intensities; the peak-normalised 0-1 vibronic PL peak in the $PL_{\perp}$ spectrum shows a larger intensity than that on the corresponding $PL_{//}$ spectrum, indicating that the $PL_{//}$ and $PL_{\perp}$ emission from the aligned F8BT at location #4 originate from two different emissive excited electronic states whose radiative decay can be polarised along two directions. The $PL_{//}$ spectrum is dictated by intrachain PL emission that is shown to linearly polarize along the direction of the F8BT transition dipole moment, whereas the $PL_{\perp}$ spectrum is formed mainly due to contributions of the emissive excited interchain species with its PL polarization direction being perpendicular to the chain alignment direction (i.e., along the interchain direction).[41] The former indication is helpful to explain the presence of a projected 0-0 vibronic PL transition band in the perpendicular collected ($\theta_{Em}$ = 90°) PL spectrum of the full-aligned F8BT monodomain *Film II* (Fig. 3e), since in this case the macroscopic emission was selected with an orthogonal polarisation angle relative to the chain alignment direction, which is still at an deviation angle of ~20° from the direction of transition dipole moment of the photoaligned F8BT chains in the extended nematic monodomain. The polarization-dependent separation phenomenon between the two orthogonal polarised PL components is regarded as a result of the anisotropic structural ordering preserved in the SD1-aligned F8BT nematic monodomain. In particular, the additional interchain emission identified in the highly oriented F8BT monodomain is expected to be structurally allowed by the remarkably elongated interchain conjugation length and the associated formation of long-range BT-to-BT electronic delocalization and strong interchain coupling. However, the minor polymeric nanocrystals dispersed in the amorphous chromophoric host in the nonaligned nematic polydomain *Film I* would trigger isotropic FRET, 3-D excitation energy concentration and no additional polarised emissive interchain species.

Another important finding from Fig. 4d-f is that the domain boundary regions exhibit the best resolved 0-2 vibronic transition peaking and the greatest $P_{(0\text{-}2)/(0\text{-}0)}$ ratio at ~580 nm, both of which can be potentially ascribed to enhancement in interchain FRET funnelling and structural order of F8BT nanocrystals. The domain boundaries that serve to interface adjacent nematic domains accommodate a higher extent of structural disorder than the domain interiors in order to smooth the heterogeneity between adjacent domains and, thus, are believed to incorporate a larger fraction of amorphous-phase F8BT (or a decreased inclusion of polymer nanocrystals). Such a low accommodation of the acceptor elements can be attributed to a more favorable situation in which each F8BT nanocrystal acceptor couples with relatively more peripheral donor elements. By assuming that the donor-to-acceptor FRET process in the self-doped host-guest system of mesophase F8BT is more efficient to induce the lower-energy 0-2 vibronic PL transition band than the 0-0 transition, a tendency of incorporating a higher level of disordered F8BT would result in a larger $P_{(0\text{-}2)/(0\text{-}0)}$ value. A similar change trend is observed in Fig. 4g-i among the nematic polydomains in a nonaligned 80 nm-thick nematic polydomain *Film I*, demonstrating a larger yet uniform $P_{(0\text{-}2)/(0\text{-}0)}$ and a slightly blue-shifted 0-2 vibronic μ-PL peak relative to that of the nonaligned 160 nm-thick polydomain background. In other words, additional PL contribution from an increased $P_{(0\text{-}2)/(0\text{-}0)}$ ratio may correlate with the occurrence of more efficient interchain exciton migration owing to lower energy losses over a shorter interchain distance before the radiative decay. Accordingly, improved PLQE up to 70% ~ 72% is determined in various nonaligned but thinner (80 nm-thick) F8BT nematic polydomain *Film I*; a reduction in film thickness results in a smaller spatially-averaged domain size across the nematic polydomains and, thus, a slightly larger fraction of domain boundary regions and red-shifted absorption spectra would correspond to larger spectral overlapping for more efficient FRET (Fig. S10). Based on the minimization of the energy of domain bulk against domain boundary energy, we arrive at a power law scaling of the spatially-averaged nematic polyamine domain size ($w$) ($w \propto D^n$) with the film thickness ($D$), which has nearly linear exponent of $n \approx 1.14$ (see details in Supplementary Section III).

## 3. Conclusion and Overlook

This work systematically studies the structure-property relationship in solution-processed light-emitting LCCP films, via fine-tuning the structural ordering (order parameter ranging from 0 to ~100%) by making use of the LC-phase ordering and high-quality photoalignment of chain orientation. Bringing structural order into in the otherwise disordered F8BT films enables us to optimize the spatial distribution and interactions of physical structures and enable polarised photophysical properties. The PL polarization anisotropies in the photoaligned F8BT nematic monodomains are generally larger than the corresponding optical dichroic ratios, since only partial chain segments contribute to emission but the whole backbones are active for optical absorption. Even for a nearly 100% order parameter, as demonstrated in the photoaligned F8BT nematic monodomains,[40] the GIWAXS data indicates imperfect (herein 60% - 73%) overall crystallinity, confirming that the degree of polymer crystallinity is also sensitive to the local arrangement of side-chains and π-π stacking between backbones.

We demonstrate that the commonly-used pristine (e.g., spin-coated and nearly amorphous) conjugated copolymer layers tend to adopt exceedingly twisted chain-conformations and possess a significant population of localised interchain interactions in order to minimise steric hindrance. This disordered structural configuration is fundamentally responsible for the absence of long-range structural order, a dominance of the low-energy vibronic transition bands in the PL spectra, and the lowest PLQE therein; a high degree of structural disorder also gives rise to a short-ranged but strong electronic coupling between the excited state and the ground state and, thus, shorten PL lifetime as well. The mesophase long-range orientational ordering alone is adequate for facilitate the self-assembly of a minor fraction of polymer nanocrystals (with the most extended chain-conformations) in a pool of disordered LCCP chains. This self-doping within the Schlieren polydomains of nematic F8BT then reconciles a bioinspired host-guest system to make use of the nonradiative 3-D donor-to-acceptor FRET funnelling and eventually efficient intrachain PL emission of F8BT nanocrystals. For this self-doped host-guest system, there exists sufficient spatial difference in the energy level of polymer (semi-)crystals and that of the amorphous host, which facilitates favorable excitonic energy concentrations and therefore considerably enhances the overall PLQE and defect tolerance. The LC-phase structural reorganization in the nematic F8BT polydomain film gives rise to PLQE values >70%, which approaches that of the theoretical upper limit of isolated chains in the solution of F8BT. The presented light-harvesting and PLQE reinforcement mechanism (engineering the physical structures of a LCCP) is somewhat analogous to the adaption of photosynthetic organisms to fluctuating sunlight via regulating the involvement of active smaller- and larger-sized LHCs.

Macroscopic LC domain texture and relative weighting of LCCP nanocrystals in a nematic polydomain glass films play a key role in limiting the excitation energy transfer and emission efficiencies. Our polarized PL spectral mapping highlights a similar lineshape that exists between the nematic polydomains and (the $PL_{//}$ spectra of) the photoaligned monodomains. The domain boundary regions, where a greater fraction of the amorphous-phase donor is accommodated to increase the donor/acceptor spectral overlapping, can be more favorable than the domain interior when it comes to enhancing FRET/PL efficiencies and suppressing the nonradiative decays associated with traps. Although the localised polymer nanocrystals take up a low weight fraction (4% ~5%) in the nonaligned nematic polydomain film, they govern the emission properties of the F8BT glass films by making the major amorphous matrix non-emissive. Additional evidence for the creation of a self-doped host-guest system was provided by the largest PLQE within the nonaligned F8BT nematic polydomain *Film I* among all F8BT films measured in this study, and that such a high PLQE is further augmented by reducing the spatially averaged domain size of self-doped nematic polydomains. High-quality chain orientation in the photoaligned F8BT monodomain films increases optical absorption strength along the chain alignment direction and generates exceptional anisotropies in the structural order and PL polarization anisotropy. The anisotropic in-plane packing of the extended F8BT monodomain allows for an additional luminescent interchain species that exhibits a PLQE value 7 times lower than that of the intrachain emission;

these structural and coupling aspects are basis for a linear decrease of PLQE with increasing chain-alignment quality. Strong or long-range electronic coupling, especially when involving *H*-aggregation and close acceptor-to-acceptor contacts between adjacent copolymer chains, is accompanied by a considerably low PLQE because in this case the interchain energy transfer is vulnerable to traps with a tendency of arriving at red-shifted charge-transfer states.[42] A substantial enhancement in polymer crystallinity of the highly-oriented, large-area-extended F8BT nematic monodomains gives rise to polarisation-dependent PL spectral separation of the two orthogonal emission components of distinct characters that are not possible in non-aligned layers, alongside the broadened PLE spectra pointing to access to extra band-edge and sub-band states. An interplay of mesophase self-assembly and SD1 molecular templating effect in a LCCP glass film makes it possible to access the most localised electronic states in the vicinity of strong many-body interaction regimes.[43]

Topics for further study include: *i*) Use of ultrafast two-dimensional spectroscopic methods (e.g., 2-D electronic-vibrational spectroscopy[44-46]) to separate intrachain and interchain energy transfer pathways in self-doped and photoaligned LCCP layers and to study the dynamics and coherence of involved electronic or vibrational coupling. *ii*) Application of spatial patterning of high-quality chain orientation into other LC and LC/molecular hybrid systems to unlock the full potential of molecular alignment and extended polymer (semi-)crystals, such as inscribing bespoke in-plane patterns for refractive index modification, which may transform the design and manufacture of planar optical cavities, optical circuits, and meta-structures, alongside wiring with natural photosystems towards semi-artificial photosynthesis and quantum coherent excitation energy funnelling.[46,47] *iii*) Combining the LC-alignment and self-doping approach explored in the present study with external doping methods[48,49] could bring structural ordering into some high-performing bioelectronic, thermoelectric, bio-mimicking and photoelectrocatalytic organic semiconductors.[50-52]

## 4. Experimental Methods

**4.1. Materials and Film Preparation**: The photo-alignment layer material, SD1, was provided by DIC Corporation Japan and used as received. The F8BT polymer with molecular weight, $M_w$ = 55,000 and polydispersity index, PDI = 2.3 was purchased from ONE-Material Inc. The photoalignment layers were spin-coated from SD1 solutions in anhydrous 2-methoxyethanol (≥ 99.8%) onto pre-cleaned Spectrosil substrate at 500 rpm for 5 seconds then 2000 rpm for 20 seconds using SD1 solution concentrations of 0.5 mg/ml and 0.1 mg/ml for the fabrication of continuous and discontinuous SD1 layers, respectively. These discontinuous SD1 films were used to induce partial domain alignment in the overlying F8BT films in the nematic phase while the continuous SD1 layers (< 4 nm in thickness) were employed to fully align the overlying F8BT nematic films. After spin-coating, the SD1 photoalignment layers were annealed at 110 °C for 10 minutes in ambient conditions to ensure solvent removal, which was followed by molecular alignment performed via uniform polarized UV light exposure as described below. F8BT was then spin-coated onto the UV-aligned continuous or discontinuous SD1 layers using the same filtered 30 mg/ml solution in anhydrous toluene. The thickness of the F8BT films were tuned by systematically varying the spin-coating speed and duration. The F8BT films were vacuum-dried for >1 hour to remove excess solvent and then subjected to thermotropic alignment as described below.

**4.2. UV Alignment of SD1 Photoalignment Layers**: SD1 photoalignment layers were aligned by irradiating the samples in air with 5 mW/cm$^2$ linearly polarized 365 nm light (close to the absorption peak of SD1) from a Thorlabs CS2010 UV-curing LED system equipped with a WP25M-UB broadband wire-grid polarizer. For spatial patterning of the alignment region in the SD1 alignment layer, polarized UV exposure was used in combination with a specially-made photomask placed above the SD1 layer (with spatially patterned silver surface facing the top surface of SD1 layer). The spatially defined photomask pattern was fabricated based on a standard photolithography technique for spatially patterning a silver-coated (150 nm thickness) glass substrate (see also the fabricated silver pattern on the photomask in Fig. S7). The duration of the polarized UV irradiation for aligning the continuous and discontinuous SD1 layers was varied over a range from 3 mins to 20 mins so as to produce varying levels of structural order in the SD1 layers which were then utilized to produce the dichroic ratios in the overlying F8BT films, as shown in Fig. 3b.

**4.3. Thermotropic Alignment of the F8BT Films by UV-aligned/patterned SD1 Layers**: Thermotropic alignment of F8BT films that have been spin-coated on top of aligned/patterned, continuous/discontinuous SD1 photoalignment layers was carried out using a Linkam THMS600 hotstage. The sample was heated at 20 °C /min to 265 °C (melt) before being slowly cooled at 3 °C/min to 250 °C (nematic phase) and then held for 10 mins. Subsequently, the sample was rapidly quenched to room temperature by transferring the sample from the hot-stage onto a copper bar. This quenching step can effectively prevent (re)crystallization and thereby "freezes-in" the SD1-aligned LC ordering of the F8BT polymer chains, yielding a highly oriented nematic glass film. Alternatively, quenching F8BT directly from its melt above the clearing temperature of ~265 °C to room temperature resulted in glassy-phase F8BT films. All steps were carried out in a nitrogen atmosphere glovebox operated with < 1 ppm moisture and oxygen content. F8BT film thicknesses were measured using a Dektak profilometer.

**4.4. Polarized Micro-PL Measurement**: Micro-PL spectra were recorded at room temperature with linearly polarized 405 nm, 150 fs, 76 MHz excitation from a frequency doubled Ti:sapphire laser, using the schematic setup shown in Section 1.5 in the Supplementary Information. The spot size at the sample was ≈ 1 μm diameter; the scanning step of the high precision *x*-*y* positioning stage was set as 200 nm in both *x* and *y* scanning directions.

## Conflicts of Interest

There are no conflicts to declare

## Acknowledgements

Y.S. gratefully acknowledges the Hong Kong Jockey Club Graduate Scholarship at the University of Oxford for full-cost financial support throughout Oxford PhD study, and additional funding from the Rank Prize Funds through the Optoelectronic Covid-19 Response Grant. Y.S. also thanks Professor Donal D. C. Bradley for project discussions. All authors collectively thank Professor Robert Taylor for allowing access to the Micro-PL facility, Dr. Ian Dobbie at the Oxford Micron Centre for Advanced Bioimaging for assistance with the confocal fluorescence imaging, as well as Dr. Bingjun Wang & Professor Moritz Riede (University of Oxford) and Dr. Thomas Derrien (Diamond Light Source) for assistance with the GIWAXS measurement.

## Appendix A. Supporting Information

Supporting Information to this article is available from the Publisher Online Library, or from the corresponding author upon reasonable request.

# Supplementary Information

## Fine-tuning the Microstructure and Photophysical Characteristics of Semiconducting Conjugated Copolymers Using Photoalignment and Liquid-crystalline Ordering

Yuping Shi,* Katharina Landfester, and Stephen M. Morris*

Dr. Y. Shi, Prof. K. Landfester
Max Planck Institute for Polymer Research, Mainz 55128, Germany
Email: shiy@mpip-mainz.mpg.de

Dr. Y. Shi, Prof. S. M. Morris
Department of Engineering Science, University of Oxford, Parks Road, Oxford, OX1 3PJ, UK
Email: stephen.morris@eng.ox.ac.uk

## Table of Contents

## SECTION I. Additional Description of Experimental Method

### 1.1 Grazing-incidence Wide-angle X-ray Scattering (GIWAXS) Characterization

GIWAXS measurements of the different types of F8BT films were carried out at the Surface and Interface Diffraction beamline (I07) at the Diamond Light Source (DLS) using a beam energy of 20 keV (0.62 Å) and a Pilatus2M area detector. The samples were probed while inside a vacuum chamber at a pressure of around $10^{-3}$ mbar with the MINERVA setup.[1] The sample-to-detector distance was 41.8 cm as determined via AgBeh calibration. Images were converted to 2D reciprocal space using the DAWN software package.[2]

### 1.2 Polarized UV-vis Absorption and Polarized PL Spectra

Polarised UV-vis absorption spectroscopy was carried out using a PerkinElmer Lambda 1050. A Glan-Thompson polarizer was mounted in front of the F8BT films to generate a linearly polarized incident beam. The relative orientation of the polymer chains to this incident light polarization was varied by rotating the F8BT film in the vertical plane on a rotation stage. Polarized PL spectra were carried out using Horiba FluoroMax-4 and acquired by mounting a Glan-Thompson polarizer in front of the F8BT films to align the excitation ($\lambda$ = 450 nm) polarization and using a build-in rotatable polarizer in the vertical plane to control the direction of PL collection polarization.

**1.3 PLQE Measurements**: Non-polarized PLQE measurements were performed the Horiba Quanta-Phi diffusely-reflecting integrating sphere attachment of Horiba FluoroMax-4 using non-polarized 450 nm excitation. Three film samples were measured for each type of F8BT films and for each kind of film sample two measurements were carried out with the second test performed by rotating the film by 90$^{o}$ in the horizontal plane relative to the first measurement so as to study the effect of polymer chain alignment direction on PLQE. For the F8BT solution sample, four measurements were carried out with different cuvette positions in the integrated sphere. The PLQE values were the calculated using the methodology detailed in Ref. [3].

### 1.4 Polarised Micro-Photoluminescence Spectral Measurements

Micro-PL (μ-PL) spectral characterisation resolves both the PL intensity and spectrum at a given scanning pixel. The μ-PL spectral maps of LCCP films presented were recorded at room temperature with linearly polarised 405 nm, 100 fs, and 76 MHz excitation from a frequency-doubled Ti: sapphire laser, using the schematic setup illustrated in Fig. S1. The excitation laser was focused on the top surface of an F8BT film by a 100× objective with a numerical aperture of $NA$ = 0.7. The spot size of the incident laser beam focused on the sample was $\approx$1 μm in diameter. The polarisation of the incident light and the collected PL spectra were controlled using a combination of linear polarisers and half-wave plates in front of the objective and the spectrometer, respectively. The excited PL emission was collected by the same objective, dispersed by a 0.3-m-long spectrometer with a 300 lines/mm grating and detected by a thermoelectrically cooled charge-coupled device (CCD). For alignment purposes, the sample was illuminated by a broadband visible light source (yellow path) and an image of the sample was then collected by a CCD camera. A 430 nm long pass filter was used to remove the excitation laser. For the mapping of the polarised μ-PL spectra, the samples were held on a piezo-electric controlled platform and a scanning step size of 200 nm in both *x* and *y* axes.

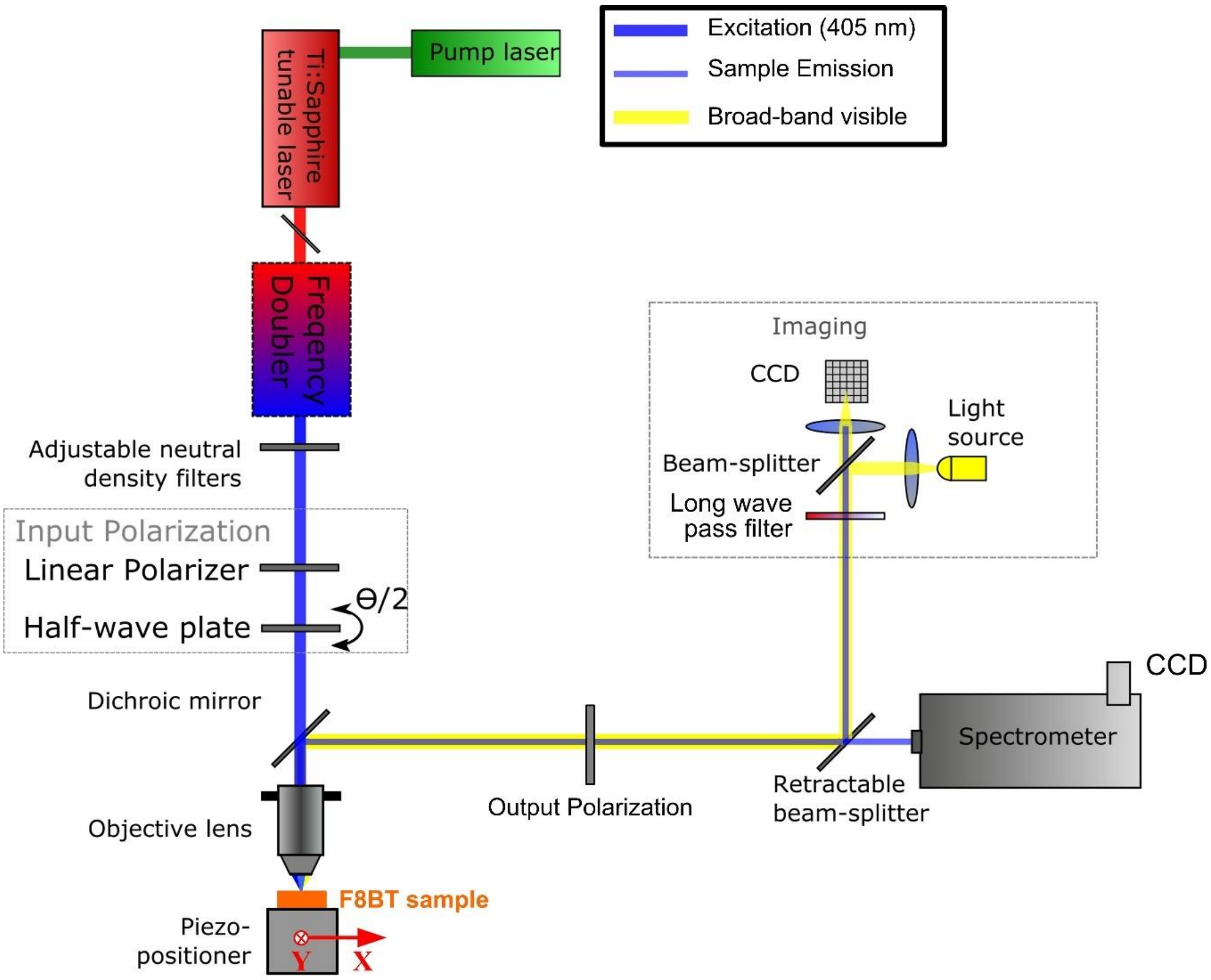

**Fig. S1** Experimental setup for the polarised μ-PL spectral mapping measurements.

## SECTION II. Supplementary Tables & Figures

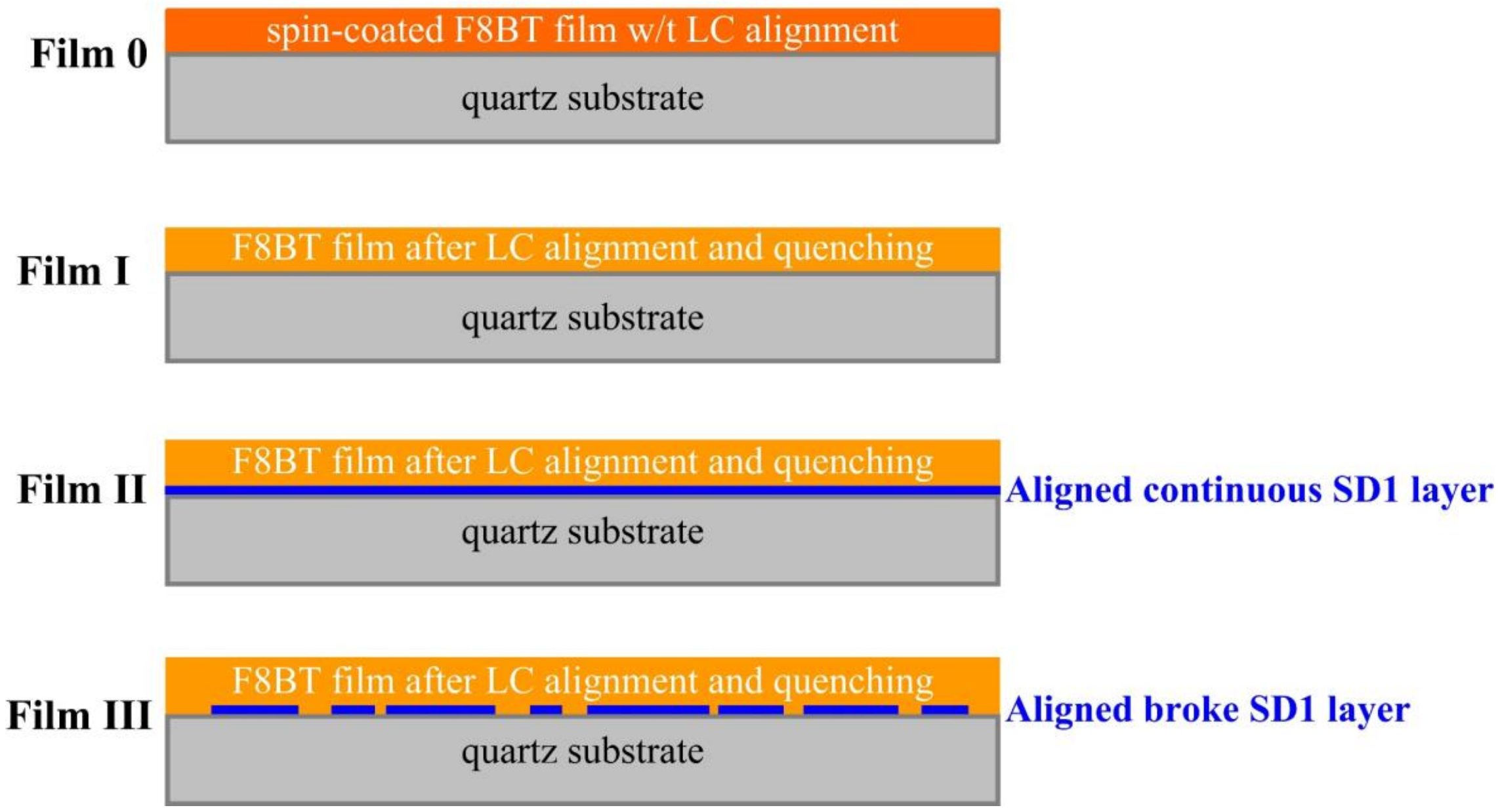

**Fig. S2** Illustration of the layer structure of the four kinds of F8BT films.

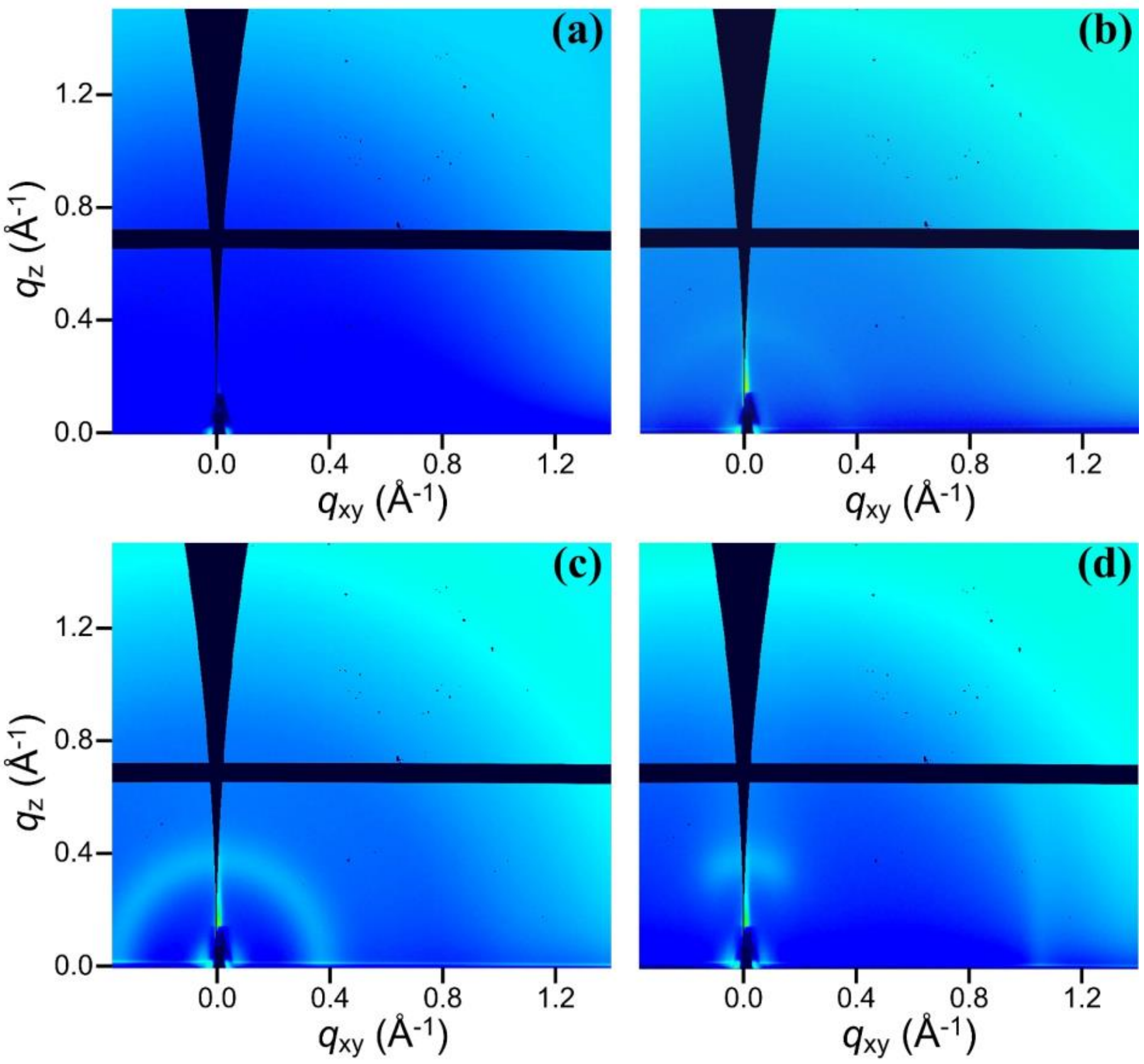

**Fig. S3** Raw GIWAXS data (i.e., before background and baseline subtraction) recorded from (a) a spin-coated non-LC F8BT film (*Film 0*), (b) a nonaligned F8BT nematic glass film (*Film I*), as well as (c - d) a fully-aligned F8BT nematic film (*Film II*) when the propagation of the incident X-ray beam is either aligned parallel (c) or perpendicular (d) to the chain alignment direction therein. All measured F8BT films are ~160 nm in thickness.

**Table S1** The *d*-spacing distances and extracted coherence lengths from the 2D GIWAXS images recorded from the nonaligned and fully-aligned F8BT nematic glass films.

| F8BT Sample | Direction | Peak | *q* (Å$^{-1}$) | *d*-spacing (Å) | Coherence length (Å) |
|---|---|---|---|---|---|
| Nonaligned | In-plane | (100) | 0.365 | 17.3 | 141.7 |
| Aligned, // | In-plane | (001) | 0.374 | 16.8 | 120 .8 |
| Aligned, ⊥ | In-plane | (010) | 1.055 | 6.0 | 205.1 |
| Nonaligned | Out-of-plane | (100) | 0.394 | 15.9 | 149.6 |
| Aligned, // | Out-of-plane | (100) | 0.385 | 16.3 | 98.4 |
| Aligned, ⊥ | Out-of-plane | (100) | 0.387 | 16.2 | 103.5 |

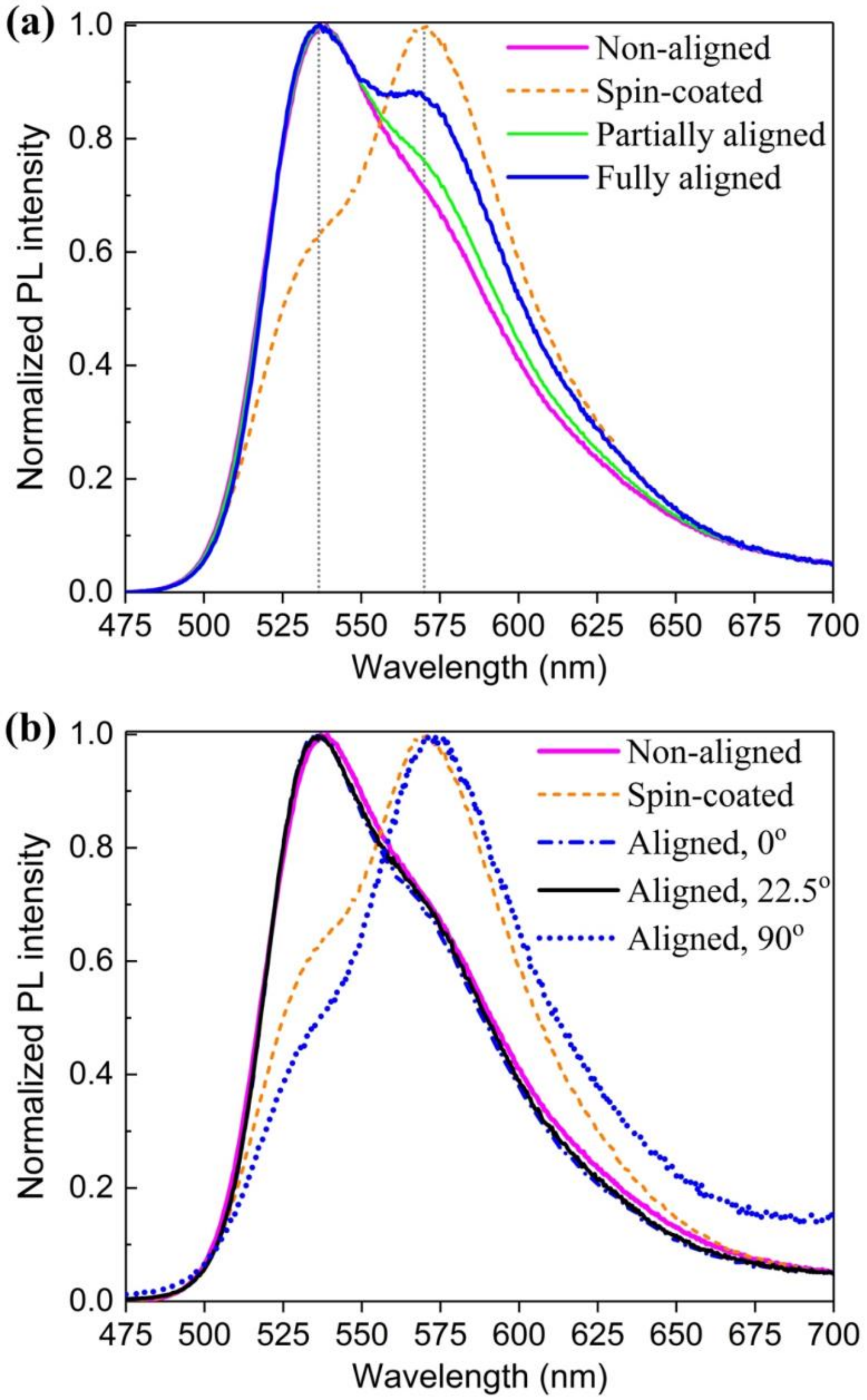

**Fig. S4** (a) Peak-normalized non-polarized PL spectra from the spin-coated non-LC F8BT film (*Film 0*), as well as the non-aligned (*Film I*), fully aligned (*Film II*) and partially aligned (*Film III*) F8BT nematic glass films. The vertical dotted lines indicate the location of the wavelength of the 0-0 and 0-1 vibronic PL peaks. All measured films herein are 160 nm thick, and PL measurements were carried out at room temperature. (b) Direct comparison of the peak-normalized non-polarized PL spectra of both the spin-coated non-LC film and the nonaligned nematic glass films with the polarised PL spectra collected from the fully-aligned F8BT nematic monodomain glass film using the three emission collection polarization angles: $\theta_{Em}$ = 0° ($PL_{//}$ spectrum), 22.5° (along the direction of optical transition dipole moment) and 90° ($PL_{\perp}$ spectrum), as labelled in Fig. 3e in the main text.

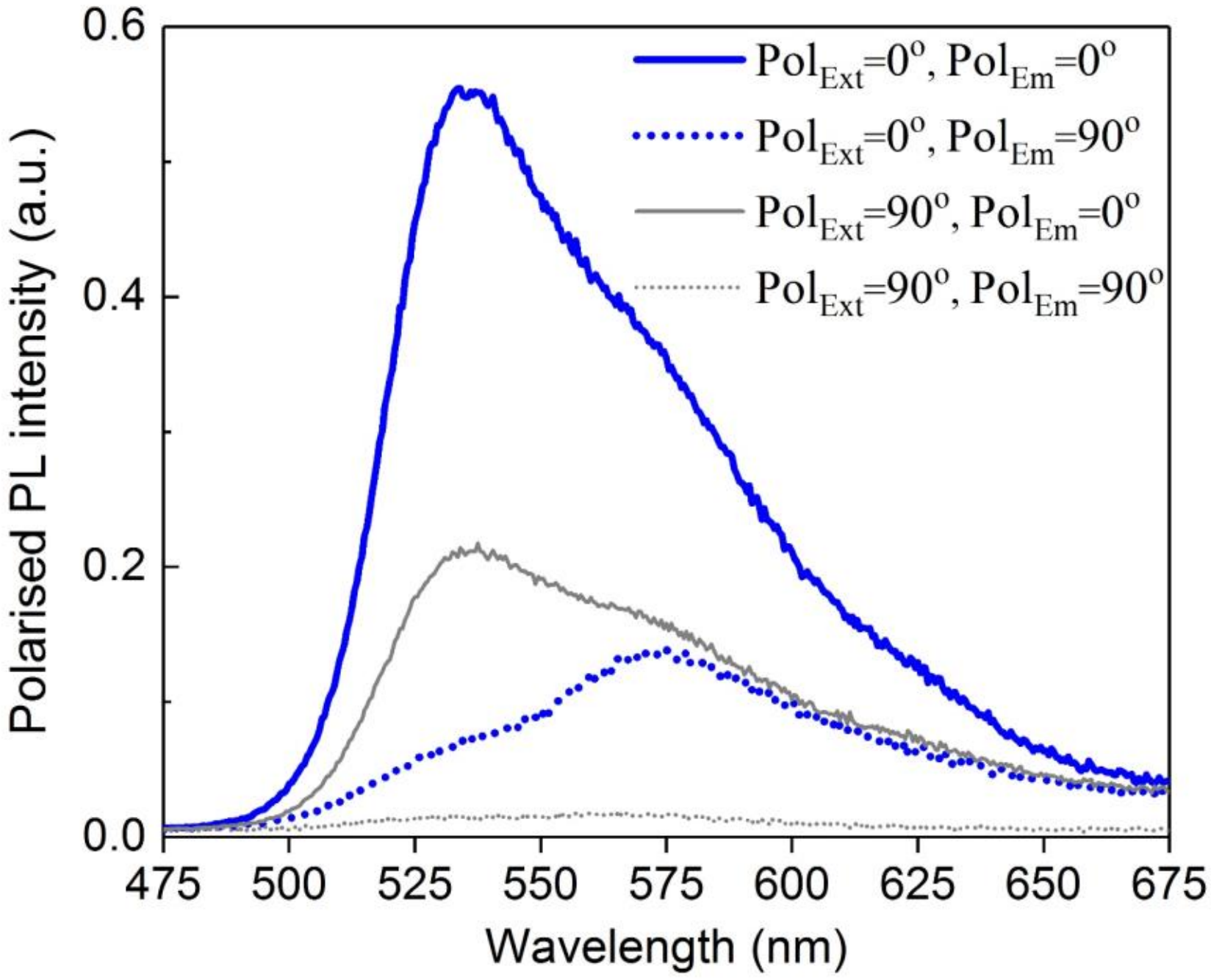

**Fig. S5** The polarised PL spectra collected from the fully-aligned F8BT nematic monodomain film (160 nm thickness) using the four combinations of the excitation polarization angle (Pol$_{Ext}$) and emission collection polarization angle (Pol$_{Em}$). Here, Pol$_{Ext}$ and Pol$_{Em}$ were defined relative to the chain alignment direction in the oriented film.

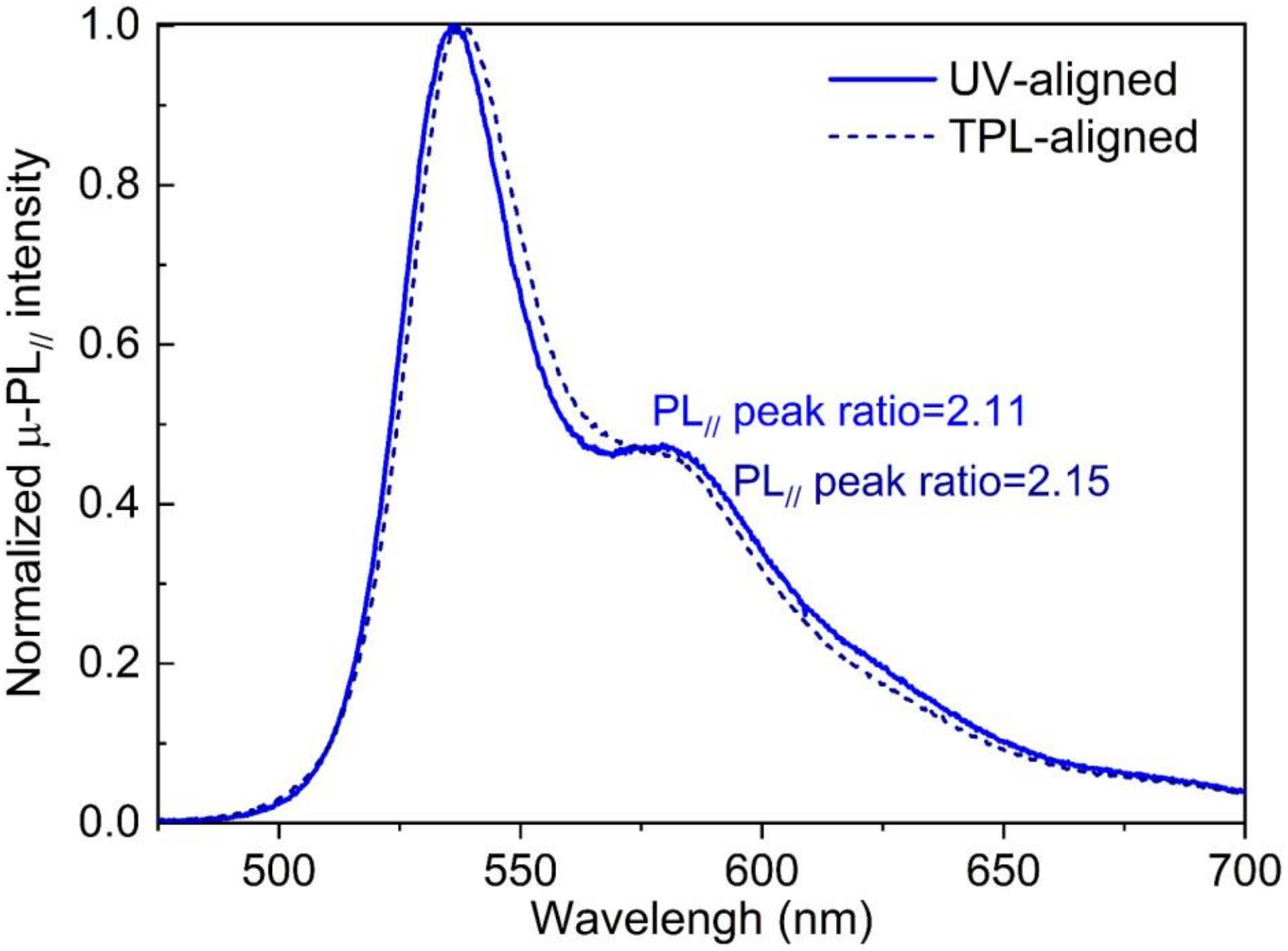

**Fig. S6** Micro PL// spectra in our best UV-aligned F8BT nematic monodomain film (film thickness = 190 nm) and also in two-photo laser (TPL) aligned nematic monodomain patter (size: 100×100 µm square). Vibronic PL peak ratios of >2 were obtained from these two µ-PL// spectra.

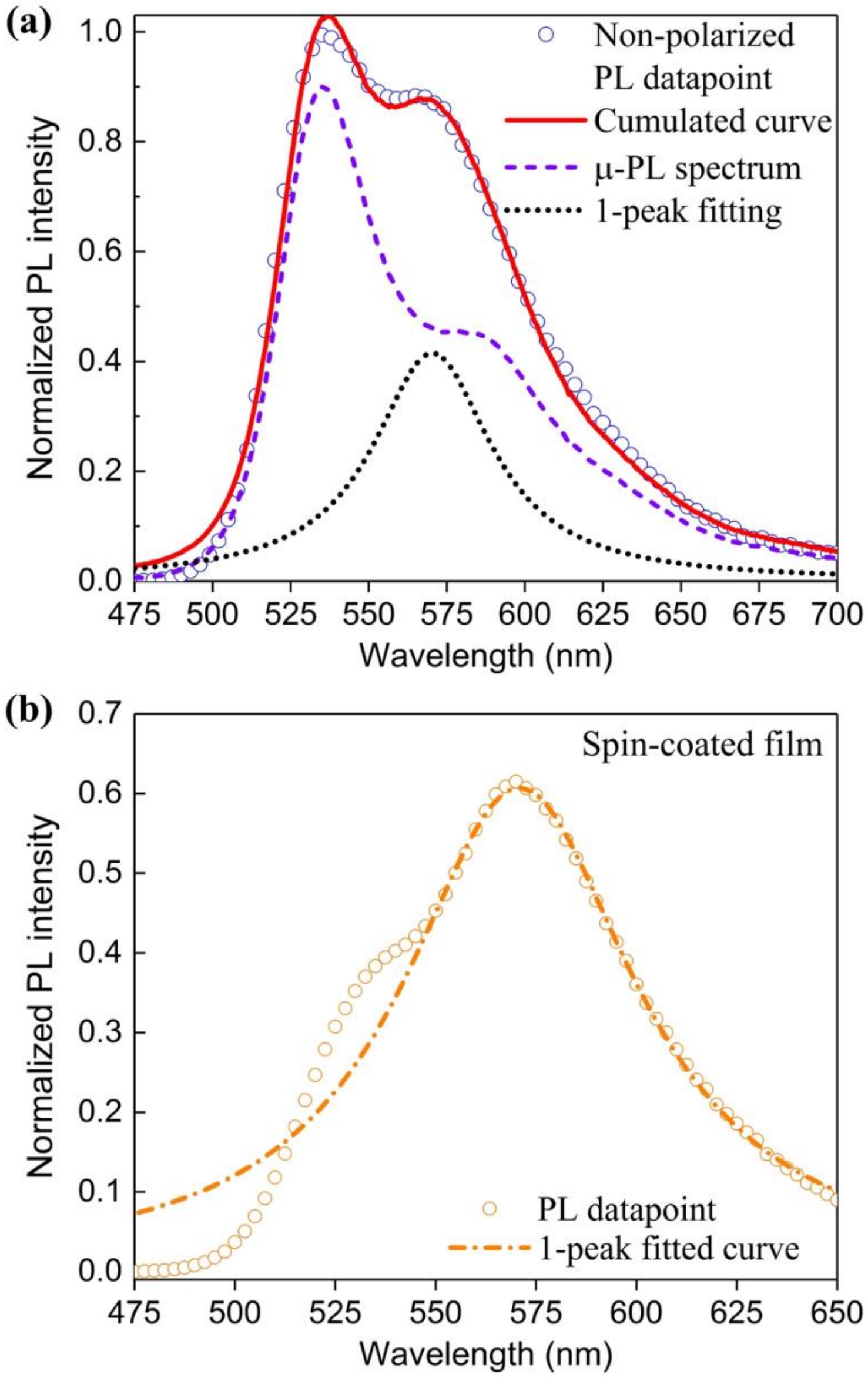

**Fig. S7** (a) Illustration of two-component fitting (the solid red curve) of the non-polarised PL spectrum (the blue open circles) collected from the fully-aligned 160 nm F8BT nematic monodomain glass film shown in Fig. 3c. While one PL component (the dotted black curve) used a Lorentzian peak centered at 570 nm with its width and peak intensity being set free during implementation of the least-square fitting, the other PL component was determined by summating the two orthogonal polarised micro-PL spectra (shown in Fig. 4d for location #4 in the photoaligned line), namely, the lineshape of the PL spectrum marked by the blue dashed curve herein is the normalised intensity of the $PL_{//}$ + $PL_{\perp}$ from the photoaligned F8BT monodomain line (Location #4) of the same film thickness, but with 20% line boarding of both $PL_{//}$ and $PL_{\perp}$ spectra as to obtain the best fitting of the non-polarised macroscopic PL spectrum. (b) One (Lorentzian)-peak fitting of the longer-wavelength optical transition band on the nonpolarised PL spectrum collected from the spin-coated non-LC reference film (*Film 0*).

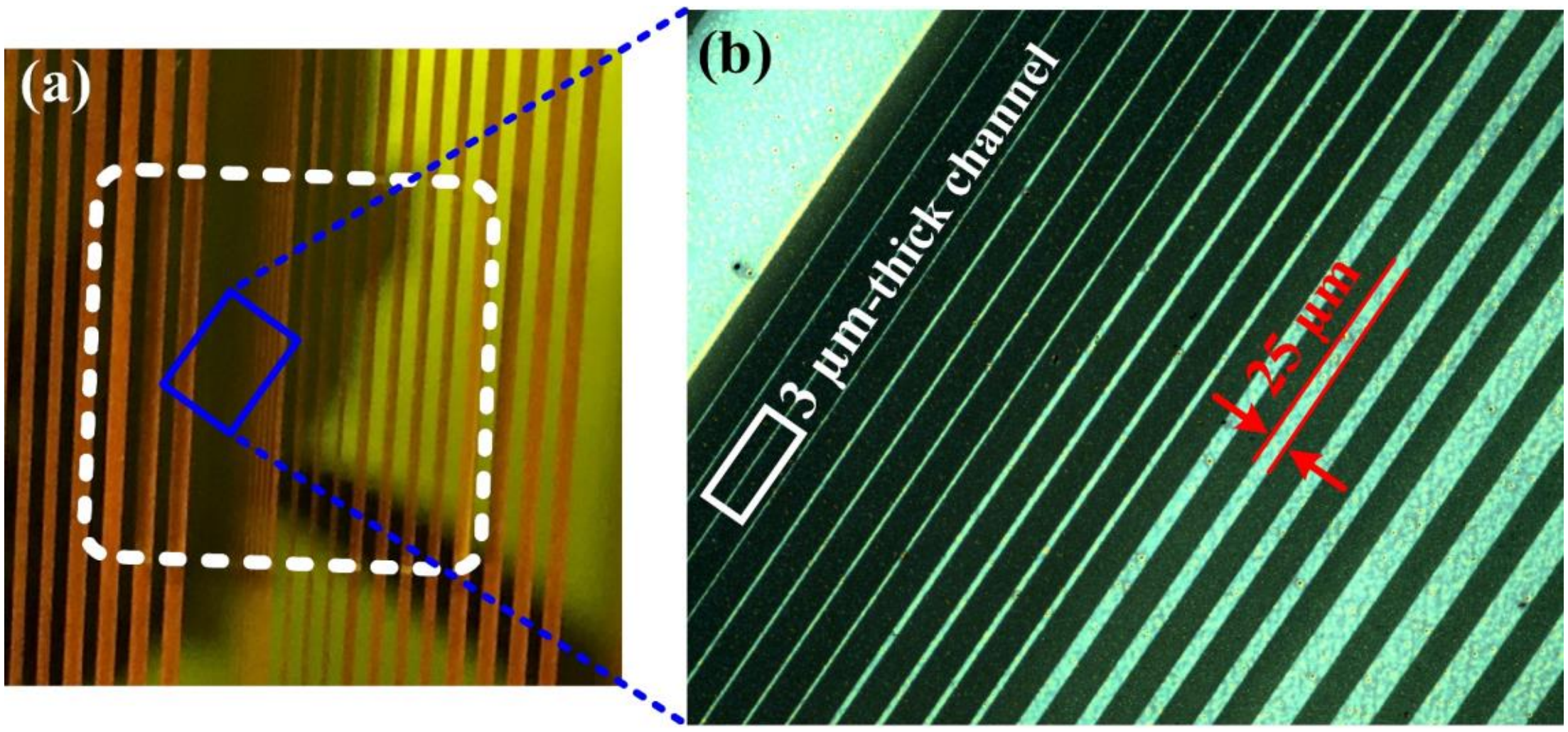

**Fig. S8** Illustration of the photo-masked UV-alignment method used to spatially define a bespoke alignment pattern in the SD1 alignment layer and an overlying F8BT glass film. (a) An optical photo demonstrating a large photomask placed on a SD1-coated quartz substrate (denoted by the dashed white square; substrate size: 25 mm × 25 mm) in the process of polarised UV illumination of the SD1 layer. (b) Bright-state polarized optical micrograph of the ensuing chain-orientation pattern in a selected region, as labelled by the blue rectangle in (a), in the thermotropically aligned and then quenched overlying F8BT glass film. In (b) there are five groups of 4 × channels with line width of 3.0 μm, 5 μm, 10 μm, 25 μm, and 50 μm; the solid white rectangle locates the UV-aligned 3 μm-thick F8BT channel against the nonaligned nematic glass background, both of which have been particularly investigated using the polarised μ-PL mapping measurements.

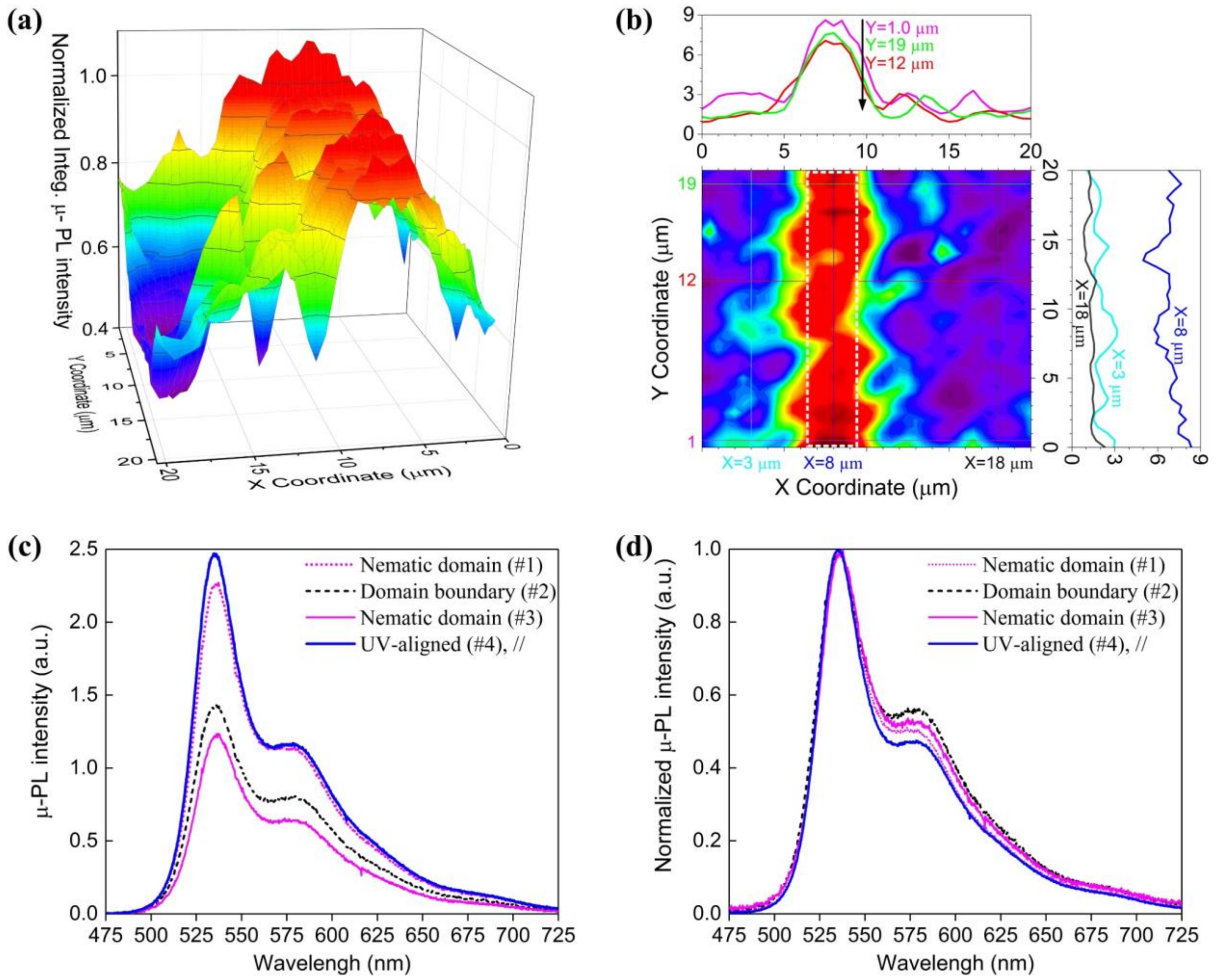

**Fig. S9** (a) 2-D μ-PL mapping of the integrated $PL_{//}$ spectral intensity for the UV-aligned line (3 μm thick) against the polydomain background. (b) Color-coded contour plot and linecut profiles of the integrated PL anisotropy ratio in the six labelled different lines along the *x*-axis and *y*-axis (i.e., $x$ = 3 μm, 8 μm, and 18 μm; $y$ = 1 μm, 2 μm, and 19 μm) in the colored-coded contour plot. The white dashed box projects the shape and location of the photomask used to pattern the UV alignment of the as-shown 3 μm-thick line in the photoalignment layer. (c) Corresponding $PL_{//}$ spectra and (d) Peak-normalized $PL_{//}$ spectra measured at the four different locations labelled in Fig. 4b in the main text. The spot #1 and #3 is located at the trough and peak PL intensity of the nematic domain, respectively, and the location #2 on the domain boundary and the location #4 representing a typical spot in the UV-aligned line.

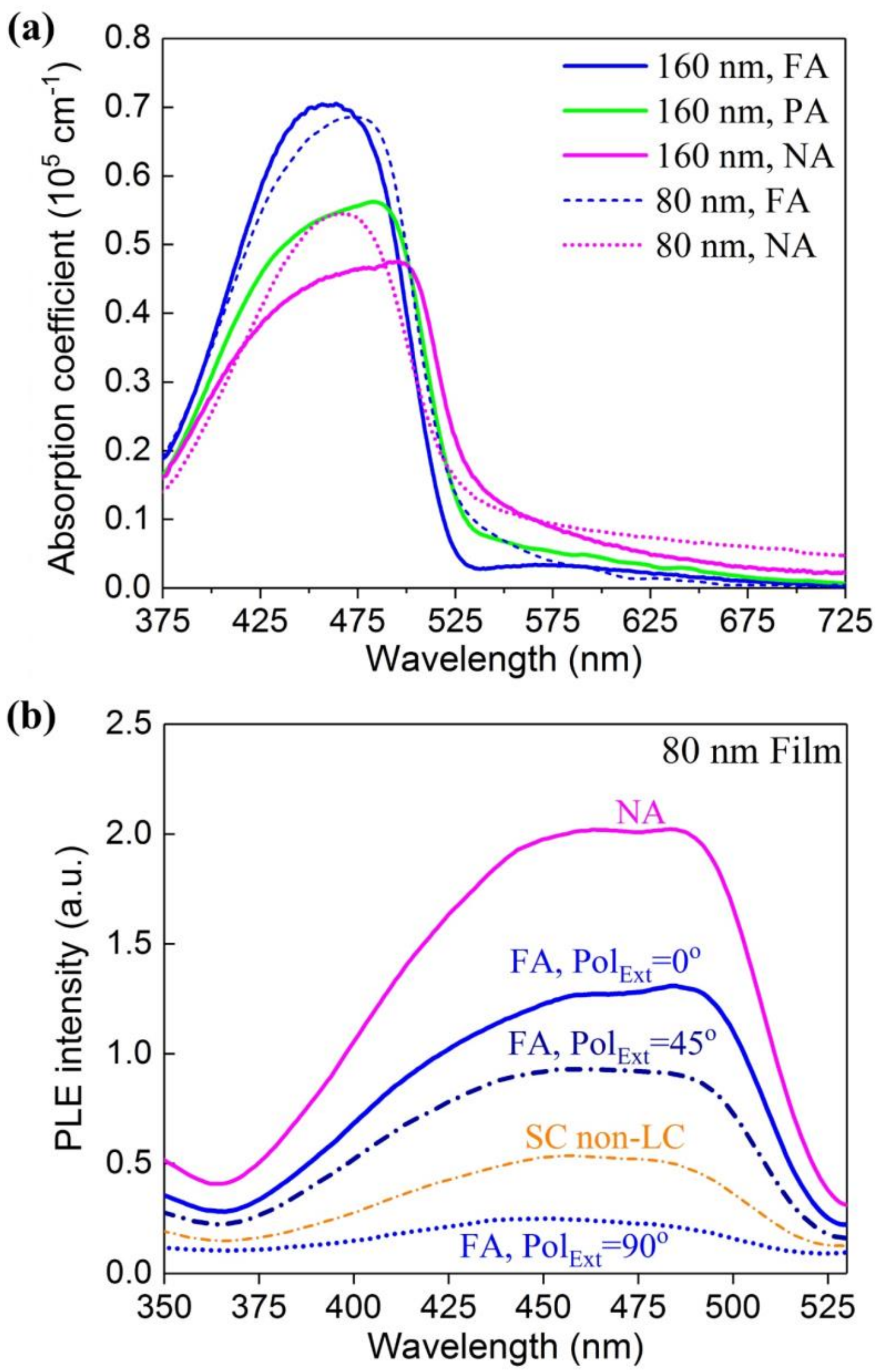

**Fig. S10** (a) Direct comparison of the absorption coefficient (i.e., the ratio between absorbance and film thickness) spectra in the nonaligned (NA, *Film I*), fully-aligned (FA, *Film II*) and partially-aligned (PA, *Film III*) F8BT nematic glass films with the thickness being 160 nm or 80 nm. (b) Polarized photoexcited ($Pol_{Ext}$ denoting the angle between the excitation polarisation and the chain alignment direction) and non-polarised collected PL excitation (PLE) spectra of the different F8BT films for emission at the 0-0 vibronic PL peak. All measured films in (b) are 80 nm in thickness. We also found that the corresponding PLE spectra for 570 nm emission (not shown here) demonstrate the same trend as these for the 0-0 vibronic PL peak herein.

## SECTION III. Domain Size Engineering and Scaling vs F8BT Film Thickness

Physical tuning of the domain boundaries in the nematic polydomain glass films complements the photo-patterning of the chain orientation as to tune the photophysical properties of the F8BT films. A deciding factor for the spatially averaged domain size and, thus, the fraction of domain boundary regions in the self-assembled F8BT nematic polydomains is the thickness of the glass films. Fig. S10 showcases the results of a demonstration of domain engineering via systematically varying the thickness of the *nonaligned* F8BT nematic glass films (*Film I*) and its merit in enhancing the PLQE. The thickness of the nonaligned F8BT nematic glass films was tuned from 40 nm to 480 nm (measured using a Dektak profilometer) by only varying the speed of the spin-coating deposition with the 30 mg/mL F8BT solution in anhydrous toluene. The colours in the crossed polarising optical microscopy (POM) images visualise the phase difference of the nonaligned F8BT glass films due to the different thicknesses. Fig. S13f demonstrates that the increase in the spatially-averaged domain size with the film thickness in the nonaligned F8BT nematic polydomain films is nearly linear. Similar size scaling trends have been reported for other LCs, e.g., the thickness dependence of the lateral size of domains in Smectic LCs[4] and defect number density in the Schlieren textures of low molar mass nematic LCs.[5]

Given the shape irregularity of the polydomain LC texture observed in the nonaligned F8BT nematic glass films, the size of the nematic domains ($w$) would scale with the film thickness ($D$) according to a power law, namely $w \propto D^{H_x/(3-H_y)}$, as a result of a minimization of the energy of the domain bulk against the domain boundary energy.[6,7] Here, $H_x$ and $H_y$ denotes the Hausdorff dimensionality of the irregular F8BT nematic domains along the $x$-axis and $y$-axis in the film plane and can be then estimated as $H_x = H_y \approx 1.6$ using the extracted coherence lengths for the nonaligned F8BT nematic film.[8] Taking all together, the scaling of the domain size in the nonaligned F8BT nematic films follows: $w \propto D^{1.14}$, and a fitting result using this power law to the dataset of spatially averaged nematic domain size is illustrated by the black dashed curve in Fig. S13f. The minor deviations in fitting the domain size for the nonaligned F8BT nematic films with thickness <80 nm are likely caused by the enhanced surface confinement effect of the reorientation of F8BT chains and altered chain-entanglement state in the nematic mesophase in these ultrathin F8BT films.[9,10] These surface limiting factors were not considered in the proposed simple energy minimization model but may result in the formation of polydomain LC textures in sufficiently thin F8BT nematic glass films.

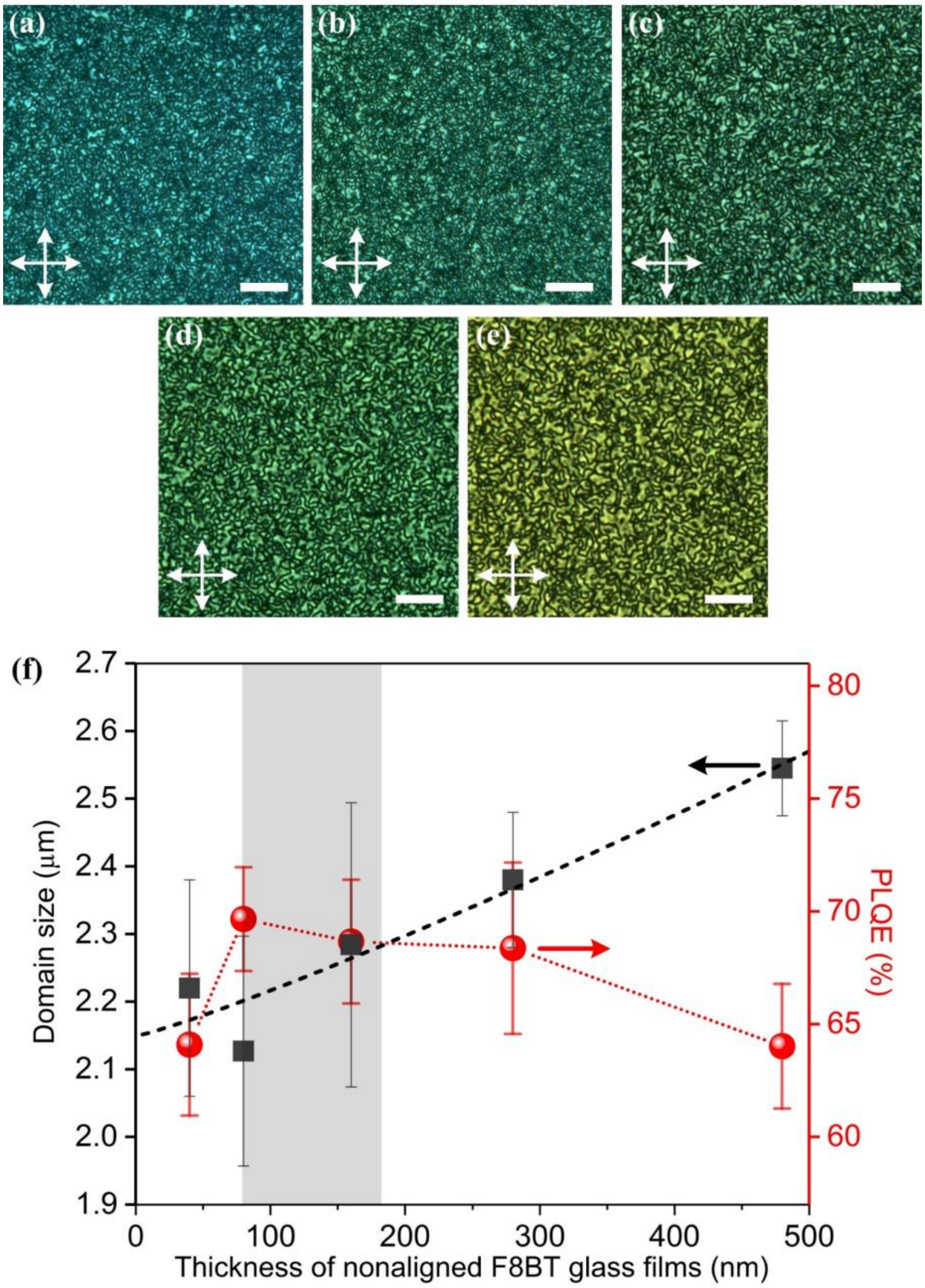

**Fig. S11** (a - e) Typical POMs demonstrating a polydomain LC texture observed in the quenched nonaligned F8BT nematic films (*Film I*) coated on a quartz substrate with a film thickness of 40 nm (a), 80 nm (b), 160 nm (c), 280 nm (d) and 480 nm (e), viewed between crossed polarisers as indicated by the two crossed white double-headed arrows. The scale bars in (a - e) are 20 µm; the colour change from (a) to (e) corresponds to an increase in retardation in the thicker films. (f) Plots of both the spatially averaged domain size and PLQE as a function of the thickness of the nonaligned nematic F8BT films. The black dashed curve represents the fitting result using the power-law scaling to the dataset for the nematic domain size. The error bars denote a variation in the corresponding data determined from four nonaligned F8BT nematic samples for each film thickness. The shaded rectangle highlights the varying range of the backbone length of the used F8BT.

## SECTION IV. Photoalignment of F8BT Nematic Films with Varying Thickness

The F8BT chain-orientation quality and LC domain pattern can be tuned through tailoring the thickness of the SD1-aligned F8BT nematic glass films that were oriented by a UV-aligned (for 5 mins) continuous SD1 layer (deposited from 0.5 mg/ml SD1 solution) before quenching the SD1/F8BT bilayers to form a solid-state glass film. The thickness of the overlying F8BT nematic glass films was engineered from 40 nm to 480 nm (measured using a Dektak profilometer) by only varying the speed of the spin-coating deposition with the 30 mg/mL F8BT solution in anhydrous toluene. The thickness dependence of the F8BT film dichroic ratio and the LC textures of these F8BT glass films are shown in Fig. S11a. It is evident that polymer chain orientation arising from the same UV-aligned SD1 alignment layers is favoured for the F8BT glass films with thicknesses in the range ~100 to 300 nm, in terms of >8 dichroic ratios. The maximum dichroic ratio is 12.3 at a film thickness of ~190 nm, which reaches the theoretical upper-limit of dichroic ratio (*DR*) by considering a deviation angle of 20° - 22° between F8BT chain axis and the transition dipole moment. Although a further increase in F8BT film thickness can constantly lower the overall F8BT alignment quality averaged across the whole thickness of the glass films, it reassuringly demonstrates that the best chain-orientation occurs for film thicknesses similar to those commonly used in a rich range of device structures.[11-13] This optimised optical dichroism for the photoaligned nematic F8BT in the solid glass nematic state is larger than the DR values reported for the oriented F8BT films by rubbed high-temperature PI alignment layer (*DR* = 8.2)[14] and by surface confinement (*DR* = 7),[15] as well as for the oriented F8BT nanofibers by an electrospinning process (*DR* = 2 -3).[16]

The polymer chains in the mesophase behave in a cooperative manner with the SD1 commanding layer acting to direct their long-range orientational ordering, most desirably into an extended monodomain state. It is expected that the interplay between chain-chain and chain-SD1 interactions facilitates the polymer chain ordering. As for a thicker F8BT film, the aligning effect/force of the optimised SD1 commanding layer will decrease with an increasing distance between the polymer chains and the alignment surface; this tends to magnify the effect of the self-organisation in the nematic phase before quenching. This could explain the emergence of the non-aligned or not-well-aligned regions in the LC textures of the 280 nm-thick Film #4 and 480 nm-thick Film #5 shown in Fig. S11b and c. On the other hand, the nematic mesophase polymer chains in a sufficiently thin (e.g., 40 nm - 80 nm in thickness) F8BT glass film may experience a different yet strong surface effect during the SD1-orienting process. This is due to the significantly reduced separation distance between the SD1 alignment surface and the free surface of the nematic F8BT, which would limit the re-orientation of the polymer chains[17,18] and result in the formation of not-well aligned regions and/or a polydomain texture.

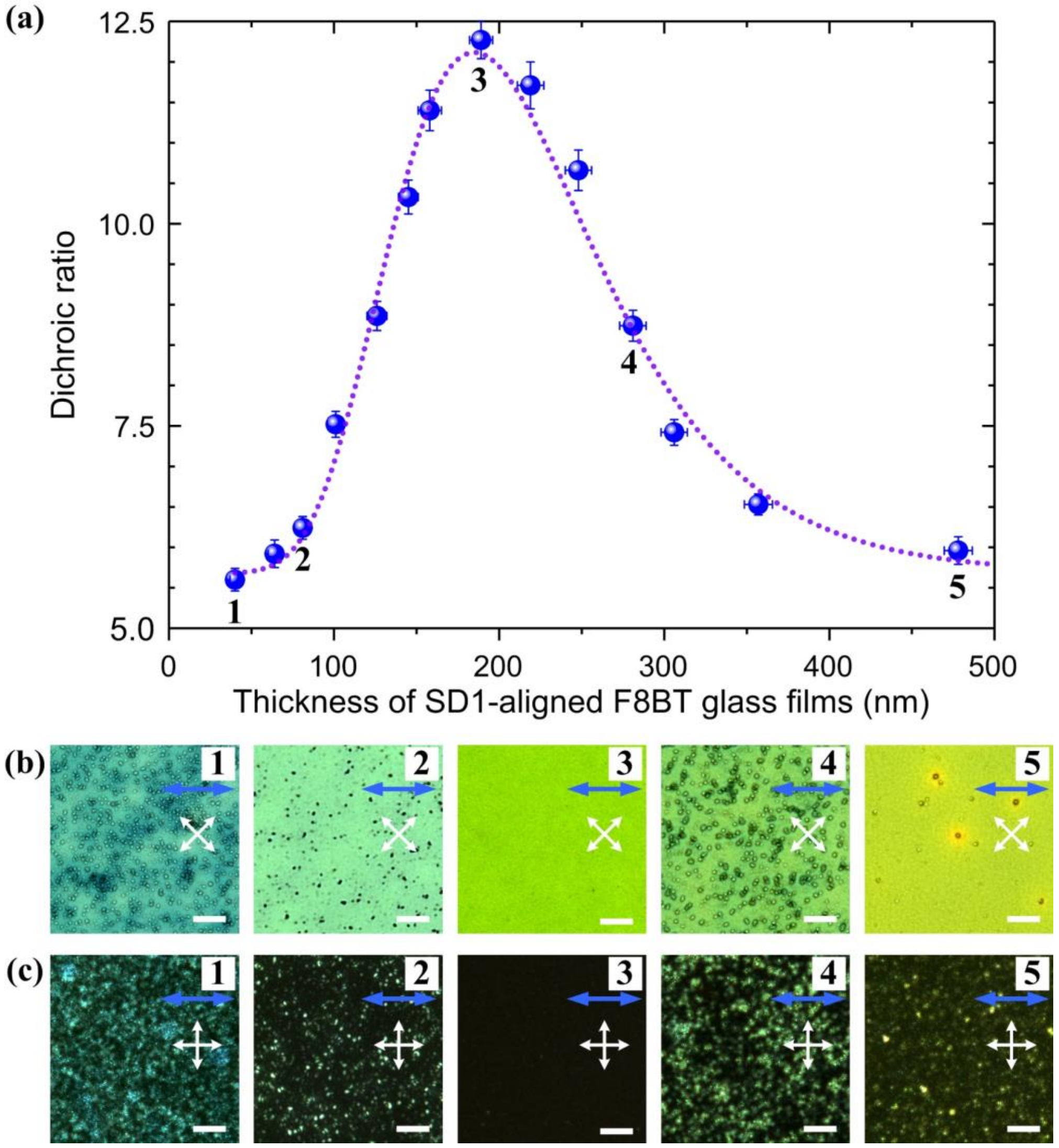

**Fig. S12** (a) Dichroic ratio in the SD1-aligned (by the UV-illuminated continuous SD1 photoalignment layers) F8BT nematic glass film as a function of F8BT film thickness. The dotted curve is a guide for the eye. (b) Bright-state and (c) Dark-state POM images illustrating the LC textures observed in the SD1-oriented F8BT nematic films with one of the F8BT film thicknesses at the five points labelled in (a), that is, from left to right: 40 nm (1), 80 nm (2), 190 nm (3), 280 nm (4) and 480 nm (5). The horizontal scale bar in each image in (b) and (c) is 20 µm. The oriented F8BT samples (chain orientation denoted by the thick double-headed blue arrows) were placed between a crossed polarised polariser/analyser pair (the crossed thin white arrows). The error bars denote a variation in the corresponding data determined from four samples for each type of F8BT film. The shaded rectangle highlights the varying range of the backbone length of the used F8BT co-polymer. The dark (bright)-state POM images were recorded using the same imaging settings. The colour change from (1) to (5) in (b) corresponds to the increase in retardation in the thicker films.